\documentclass[11pt,leqno]{article}
\usepackage{amssymb,latexsym,amsmath,bm,amsfonts,amsthm,enumerate}
\usepackage{cite}
\usepackage{hyperref}
\hypersetup{
    colorlinks,
    citecolor=black,
    filecolor=black,
    linkcolor=blue,
    urlcolor=black
}
\usepackage{helvet}
\usepackage{graphics,epsfig,wrapfig,subcaption}
\usepackage{epstopdf}
\usepackage[all]{xy}
\usepackage[rightcaption]{sidecap}

\usepackage[margin=1in]{geometry}

\def\\{\itemize}

\newcommand{\ttt}{\texttt}

\usepackage{ulem}
\usepackage[dvipsnames]{xcolor}

\newcommand{\red}{\textcolor{red}}
\newcommand{\blue}{\textcolor{blue}}

\usepackage{lineno}

\title{Online reactions to the 2017 `Unite the Right' rally in Charlottesville: Measuring polarization in Twitter networks using media followership}
\author{Joseph H. Tien\thanks{Corresponding author.  \ttt{tien.20@osu.edu}}, Marisa C. Eisenberg, Sarah T. Cherng, Mason A. Porter}

\begin{document}
\maketitle



\section*{Affiliations}

\begin{itemize}
\item{Joseph H. Tien: Department of Mathematics and Mathematical Biosciences Institute; The Ohio State University}
\item{Marisa C. Eisenberg: Department of Epidemiology, Center for the Study of Complex Systems, and Department of Mathematics; University of Michigan}
\item{Sarah T. Cherng: Precision Health Enterprise and Institute for Next Generation Healthcare, Mount Sinai Health System, New York, NY}
\item{Mason A. Porter: Department of Mathematics; UCLA}
\end{itemize}



\newpage


\begin{abstract}
Network analysis of social media provides an important new lens on politics, communication, and their interactions. This lens is particularly prominent in fast-moving events, such as conversations and action in political rallies and the use of social media by extremist groups to spread their message. We study the Twitter conversation following the August 2017 `Unite the Right' rally in Charlottesville, Virginia, USA using tools from network analysis and data science. We use media followership on Twitter and principal component analysis (PCA) to compute a `Left'/`Right' media score on a one-dimensional axis to characterize Twitter accounts. We then use these scores, in concert with retweet relationships, to examine the structure of a retweet network of approximately 300,000 accounts that communicated with the \#Charlottesville hashtag. The retweet network is sharply polarized, with an assortativity 
coefficient of $0.8$ with respect to the sign of the media PCA score. Community detection using two approaches, a Louvain method and InfoMap, yields communities that tend to be homogeneous in terms of Left/Right node composition. We also examine centrality measures and find that hyperlink-induced topic search (HITS) identifies many more hubs on the Left than on the Right. When comparing tweet content, we find that tweets about `Trump' were widespread in both the Left and Right, although the accompanying language (i.e., critical on the Left, but supportive on the Right) was unsurprisingly different. Nodes with large degrees in communities on the Left include accounts that are associated with disparate areas, including activism, business, arts and entertainment, media, and politics. Support of Donald Trump was a common thread among the Right communities, connecting communities with accounts that reference white-supremacist hate symbols, communities with influential personalities in the alt-right, and the largest Right community (which includes the Twitter account \ttt{FoxNews}).

\medskip
\noindent {\bf Keywords:} United States politics, political extremism, media polarization, social media, Twitter, community structure, principal component analysis

\end{abstract}


\section{Introduction} \label{sec1}

On 11--12 August 2017, a `Unite the Right' rally was held in Charlottesville, Virginia, USA in the context of the removal of Confederate monuments from nearby Emancipation Park. Attendees at the rally included members of the `alt-right', white supremacists, Neo-Nazis, and members of other far-right extremist groups \cite{fausset2017}. At this rally, there were violent clashes between protesters and counter-protesters. A prominent event amidst these clashes was the death of Heather Heyer when a rally attendee rammed his car into a crowd of counter-protesters \cite{duggan2019}. In the aftermath, President Donald Trump stated that there were `very fine people on both sides' \cite{shear2017}. White supremacists were galvanized by Trump's response, with one former leader stating that the president's comments marked ``the most important day in the White nationalist movement'' (\cite{daniels2018}, p.~61). Reactions to the removal of confederate statues, the violence at the rally, and President Trump's controversial response generated vigorous debate.

In this paper, we present a case study of the structure of the online conversation about Charlottesville in the days following the `Unite the Right' rally. These data allow one to study far-right extremism in the context of broader public opinion. Did support for President Trump's handling of Charlottesville extend beyond white supremacists and the `alt-right'?  Did the response to Charlottesville split simply along partisan lines, or was the reaction more nuanced? We examine these questions using tools from network analysis and data science using Twitter data from communication following the `Unite the Right' rally that include the hashtag \#Charlottesville.
Our specific objectives are to
(1) present a simple approach for characterizing Twitter accounts based on their online media preferences; 
(2) use this characterization to examine the extent of polarization in the Twitter conversation about Charlottesville; 
(3) evaluate whether key accounts were particularly influential in shaping this discussion; 
(4) identify natural groupings (in the form of network `communities') of accounts based on their Twitter interactions; and 
(5) characterize these communities in terms of their account composition and tweet content.  

Social-media platforms are important mechanisms for shaping public discourse, and data analysis of social media is a large and rapidly growing area of research \cite{tufekci2014}. It has been estimated that almost two-thirds of American adults use social media networking sites \cite{perrin2015}, with even higher usage among certain subsets of the population (such as activists \cite{tufekci2017} and college students \cite{perrin2015}). Online forums and social-media platforms are also significant mechanisms for communication, dissemination, and recruitment for various types of ethnonationalist and extremist groups \cite{daniels2018}. Twitter, in particular, has been a key platform for white-supremacist efforts to shape public discourse on race and immigration (\cite{daniels2018}, p. 64).

As a network, Twitter encompasses numerous types of relationships. It is common to analyze them individually as {\it retweet} (e.g., see \cite{conover2011,romero2011}), {\it follower} (e.g., see \cite{colleoni2014}), {\it mention} (e.g., see \cite{conover2011}) networks, and others.
There is an extensive literature on Twitter network data, and the myriad topics that have been studied using such data include political protest and social movements \cite{barbera2015,beguerisse2014,cihon2016,freelon2016,lynch2014,tremayne2014,tufekci2012}, epidemiological surveillance and monitoring of health behaviors \cite{chew2010,denecke2013,lee2014,mcneill2016,salathe2013,signorini2011,towers2015}, contagion and online content propagation \cite{lerman2010,weng2013}, identification of extremist groups \cite{benigni2017}, and ideological polarization \cite{conover2011,garimella2017,morales2015}. The combination of significance for public discourse, data accessibility, and amenability to network analysis makes it very appealing to use Twitter data for research. However, important concerns have been raised about biases in Twitter data \cite{cihon2016,mitchell2014b,morstatter2013} in general and hashtag sampling in particular \cite{tufekci2014}, and it is important to keep them in mind when interpreting the findings of both our study and others.  

There are also many studies of how the internet and social-media platforms affect public discourse \cite{bail2018,flaxman2016,sunstein2001}. In principle, social media and online news consumption have the potential to increase exposure to disparate political views \cite{wojcieszak2009}. In practice, however, they instead often serve as filter bubbles \cite{allcott2017,pariser2011} and echo chambers \cite{flaxman2016,sunstein2001}; and they thus may heighten polarization. Several previous studies have examined political homophily in Twitter networks \cite{colleoni2014,conover2011,feller2011,morales2015,pennacchiotti2011}, and some analysis has been based on tweet content and followership of political accounts \cite{colleoni2014}. We also examine political polarization using Twitter data, but we take a different approach: we focus on the homophily of media preferences on Twitter. Specifically, we examine media followership on Twitter and perform principal component analysis (PCA) \cite{jolliffe2002} to calculate a scalar measure of media preference. We then use this scalar measure to characterize accounts in our Charlottesville Twitter data set. To study homophily, we examine assortativity of this scalar quantity for accounts that are linked by one or more retweets. 
Our approach is conceptually simple, has minimal data requirements (e.g., there are no training data sets), and is straightforward to implement.  The employment of media followership is also appealing for political studies, as it is known that media preferences correlate with political affiliation \cite{budak2014,mitchell2014}.

The influence of Twitter accounts on shaping content propagation and online discourse depends on many factors, including the number of `followers' (accounts who subscribe to a given account's posts, which then appear in their feed), community structure and other aspects of network architecture \cite{weng2013}, tweet activity (and other account characteristics) \cite{barbera2015}, and specific tweet content \cite{cha2010}. One can calculate `centrality' measures \cite{newman2010} to identify important nodes in a Twitter network. There are many notions of centrality, including degree, PageRank \cite{brin1998}, betweenness \cite{freeman1977}, and hyperlink-induced text search (HITS, which allows the examination of both hubs and authorities) \cite{kleinberg1999}. In the context of our study, it is also useful to keep in mind that some structural features are particular to Twitter networks; these may influence which centrality measures are most appropriate to consider. Prominent examples of such features include asymmetry between the numbers of followers and accounts that are followed for many accounts \cite{cha2010}, automated accounts (`bots') that may retweet at very high frequencies \cite{davis2016}, and heterogeneous retweeting properties across different accounts \cite{romero2011}. The importance of such features has also led to the development of Twitter-specific centrality measures \cite{bouguessa2015,romero2011,weng2010}. In our investigation, we examine a variety of different measures of centrality for the \#Charlottesville retweet network to identify important accounts both for generating novel content and for spreading existing content.

Community detection, in which one seeks dense sets (called `communities') of nodes that are connected sparsely to other dense sets of nodes, is another approach that can give insights into network structure (especially at large scales) \cite{porter2009,fortunato2016}. Communities in a network can influence dynamical processes, such as content propagation \cite{jeub2015,salathe2010,weng2013}. Investigating community structure and other large-scale network structures can be very useful for the study of online social networks, as some accounts are anonymous and demographic data may be incomplete or of questionable validity. Community detection is helpful for discovering tightly-knit groupings of accounts that can help reveal what segments of the population are engaged in a conversation on Twitter. One can then examine such groupings, in conjunction with other tools from network analysis, to characterize communities in terms of structural network properties (e.g., distributions of degree or other centrality measures) and/or metadata (e.g., profile information), identify influential accounts within communities, and study dynamical processes on a network (such as how content propagates both within and between communities \cite{weng2013}).
 
In the present paper, we combine community detection with analysis of tweet content within and across communities. Previous studies have reported that there are differences in language in different online communities \cite{bryden2013}. Such differences can help reveal differences in demography, political affiliation, and views on specific topics \cite{colleoni2014,conover2011,wong2016}. For example, the `linguistic framing' of issues such as immigration can help reveal political orientations and agendas \cite{huber2009,lakoff2006}, and changes in language over time can reflect political movements and influence campaigns \cite{morgan2017}.  e combine community detection with tweet content analysis to compare subsets of the Twitter population who participated in the \#Charlottesville conversation by characterizing them based on the language in different communities for describing both the broader conversation topic (namely, \#Charlottesville) and specific subtopics (e.g., `Trump').

Our paper proceeds as follows. In Section \ref{sec2}, we briefly discuss our Twitter data collection and cleaning. In Section \ref{sec3}, we discuss how we characterize nodes based on media preference and PCA. In particular, we show that the first principal component provides a good classification of nodes as `Left' (specifically, nodes with a negative media-preference score) or `Right' (specifically, nodes with a positive media-preference score) in terms of their media preference. In Section \ref{sec4}, we examine the structure --- in terms of both centrality measures and large-scale community structure --- of a network of retweet relationships that we construct from our Twitter data.  
We compare central nodes on the Left and Right, and we also examine Left/Right  
node composition within communities in the retweet network. In Section \ref{sec5}, we examine the media-preference assortativity of nodes in the retweet network. Our results in Sections \ref{sec4} and \ref{sec5} allow us to gauge the extent to which the Twitter conversation, with respect to who retweets whom, splits according to media preferences. In Section \ref{sec6}, we illustrate differences in tweet content between the Left and Right communities.  Together, Sections \ref{sec4}--\ref{sec6} illustrate the extent of polarization in the Twitter conversation about \#Charlottesville. In Section \ref{sec7}, we conclude and discuss our results.


\section{Data collection}  \label{sec2}

We collected Tweets with the hashtag \#Charlottesville and the follower lists for $13$ media organizations using Twitter's API (application program interface) and the Python package \ttt{tweepy}. Public data accessibility through Twitter's API has greatly facilitated investigations of Twitter data, but such data have important limitations \cite{cihon2016,tufekci2014}, including potential biases due to Twitter's proprietary API sampling scheme \cite{cihon2016}. For example, Morstatter et al. \cite{morstatter2013} illustrated that Twitter's API can produce artifacts in topical tweet volume, potentially resulting in misleading changes in the number of tweets on a given topic over time. In our investigation, we do not consider changes in tweet volume over time; instead, we examine features of the data after aggregating over a collection-time window.

 T\"{u}fekci \cite{tufekci2014} discussed several potential issues with hashtag sampling, including different hashtag usages across different groups and discontinuation of a given hashtag once the corresponding topic has been established. (This latter phenomenon is called `hashtag drift' \cite{salganik_book}.)
   We collected the tweets that we study from a six-day period from shortly after the `Unite the Right' rally; this should lessen the potential for hashtag drift.  
  As was pointed out by T\"{u}fekci \cite{tufekci2014}, hashtag sampling draws from accounts that choose to tweet a given hashtag, and this necessarily entails biases. Nevertheless, hashtag sampling is able to provide valuable insights on the shape of online conversations.  For example, we use the collected data to examine what types of accounts chose to post tweets about \#Charlottesville. It is known, for example, that the extent that `peripheral' accounts engage in online conversations about social protest can be an important factor for content propagation on Twitter \cite{barbera2015}.

Our data collection is in accord with the Twitter Terms of Service and Developer Agreement. To protect user privacy, we include account names (i.e., ``handles'') only for Twitter-verified accounts and Twitter accounts that belong to organizations.  As described by Twitter, ``an account may be verified if it is determined to be an account of public interest" \cite{twitter_ver}. We have posted network data (without account names or tweet content), together with code for analyzing the structure of these data, at \url{https://osf.io/487fw/}.


\subsection{Tweets about \#Charlottesville}

We used Twitter's search API to sample 486,894 publicly available tweets that include the hashtag \#Charlottesville and were posted by 270,975 unique accounts between 16 August 2017 and 21 August 2017. Our data includes account name (i.e., ``handle''), time and date in coordinated universal time (UTC), and tweet content. In UTC, the earliest tweet date is 2017-08-16 22:16:21, and the latest tweet date is 2017-08-20 01:48:00. We performed our data acquisition using the Python package \ttt{tweepy}.


\subsection{Media followership}
\label{sect:media}

In December 2016, we used the Twitter API to acquire the complete lists of Twitter users who follow the following 13 media accounts: \ttt{BreitbartNews}, \ttt{DRUDGE\_REPORT}, \ttt{FiveThirtyEight}, \ttt{FoxNews}, \ttt{MotherJones}, \ttt{NPR}, \ttt{NRO}\footnote{\ttt{NRO} is the Twitter account for \emph{The National Review}.}, \ttt{WSJ}\footnote{\ttt{WSJ} is the Twitter account for \emph{The Wall Street Journal}.}, \ttt{csmonitor}, \ttt{dailykos}, \ttt{theblaze}, \ttt{thenation}, and \ttt{washingtonpost}.  At the time of access, these media accounts had significant Twitter followings, ranging from 62,078 followers (\ttt{csmonitor}) to more than 12 million followers (\ttt{WSJ}). They include both sources that studies have concluded as preferred by conservative readers and those that they have concluded as preferred by liberal ones \cite{budak2014,mitchell2014}.


\section{Twitter media preferences}  \label{sec3}

\begin{table}[ht]
\begin{centering}
\begin{tabular}{lccc}
Media account & 1st \\ \hline 
\red{\ttt{BreitbartNews}} & \red{0.4071} \\ 
\red{\ttt{DRUDGE\_REPORT}} & \red{0.3843} \\ 
\red{\ttt{FoxNews}} & \red{0.3779} \\ 
\red{\ttt{theblaze}} & \red{0.2054} \\ 
\red{\ttt{NRO}} & \red{0.0970} \\ \
\ttt{csmonitor} &$-0.0235$ \\ 
\ttt{WSJ} &$-0.1183$ \\ 
\ttt{FiveThirtyEight} &$-0.1520$ \\ 
\ttt{dailykos} &$-0.1893$ \\ 
\ttt{thenation} &$-0.2362$ \\ 
\ttt{MotherJones} &$-0.3115$ \\ 
\ttt{washingtonpost} &$-0.3321$ \\ 
\ttt{NPR} &$-0.3945$ \\ 
\end{tabular}
\caption{First principal component of the media-followership matrix $M$. The red entries designate a positive entry. For each node, we use its value in the first principal component to characterize it in the retweet network: `Left' refers to nodes with negative first principal component score, and `Right' refers to nodes with positive first principal component score.
}
\label{table:pca}
\end{centering}
\end{table}
%
%

In this section, we use media preferences on Twitter to characterize nodes in our \#Charlottesville data set.  Specifically, we find that using PCA provides an effective characterization of nodes as `Left' or `Right'. In subsequent sections, we use this characterization to examine media-preference polarization in the Twitter conversation about Charlottesville.

Of the Twitter accounts in our \#Charlottesville data set, 99,412 accounts followed at least $1$ of the $13$ media sources that we listed in Section \ref{sect:media} at the time (December 2016) that we accessed the media follower lists. Restricting to these accounts gives a 99,412\,$\times$\,13 media-choice matrix $M$ of $0$ entries (not following) and $1$ entries (following). We perform PCA on $M$, and we highlight the first component in Table \ref{table:pca}. We use the `standard' type of PCA in our investigation \cite{jolliffe2002}. For a discussion of a variant of PCA that is designed for Boolean data, see \cite{landgraf2015}.

We interpret the first component as encoding liberal versus conservative media preference, as reflected by the signs of the entries of this component. Specifically, media accounts with a positive value in the first principal component (PC) seem to correspond to accounts that previous studies have found to have a conservative slant (and to be preferred by individuals who identify as conservative), whereas accounts with a negative value in the first PC correspond predominantly to accounts that studies have concluded to have a liberal slant and/or are preferred by liberals \cite{budak2014,mitchell2014}. The sign of the score in the first PC is also consistent with conventional wisdom about liberal versus conservative leanings of these media accounts, with the exception of \textit{The Wall Street Journal} (\ttt{WSJ}), which is widely considered to be conservative-leaning \cite{flaxman2016} but has a negative first PC value. However, our findings are consistent with previous studies that, based on readership and co-citations, grouped \textit{The Wall Street Journal} with liberal media organizations \cite{flaxman2016,gentzkow2011,groseclose2005}. By contrast, previous research that examined article content identified \textit{The Wall Street Journal} as politically conservative \cite{budak2014}. Although the sign of the first PC value has a clear interpretation, the magnitude of these entries does not appear to provide an intuitive ordering (for example, with respect to a hand-curated media-bias chart \cite{mediabias4}) on the liberal--conservative spectrum.

In the rest of our paper, we focus on the value of the first PC; for simplicity, we use the term `media PCA score' to refer to this score. Positive values for this score indicate followership of the media accounts that we show in red, whereas negative values indicate followership of accounts that we show in black (see Table \ref{table:pca}). To frame our discussion, we refer to nodes with a positive media PCA score as nodes on the `Right' and to those with a negative media PCA score as ones that are on the `Left', although we note that we have not validated this measure as an indicator of political belief or affiliation. Our approach is similar to that of Bail et al. \cite{bail2018}, who applied PCA to followership of a large set of `opinion leaders' to assess political orientation.


\section{Network structure}  \label{sec4}

In this section, we explore structural features of the our {\it retweet network} $\tilde{G}$, which is a weighted, directed graph with weighted adjacency matrix $\tilde{A}$, where $\tilde{A}_{ij}$ denotes the number of times that node $j$ retweeted node $i$. In particular, we examine degree distributions (see Section \ref{sect:degree}), calculate and compare several different centrality measures (see Section \ref{sect:centrality}), and detect communities using two widely-used algorithms (see Section \ref{sect:communities}).  We combine these structural features with node characterization according to media preference (see Section \ref{sect:media}) to (1) examine how central nodes differ between Left and Right and (2) describe communities based on their Left/Right node composition.

The graph $\tilde{G}$ has 238,892 nodes, 365,589 edges, and 389,736 retweets. We focus on $G$, the largest connected component of $\tilde{G}$ when we ignore directionality (so it is $\tilde{G}$'s largest weakly connected component). The graph $G$ has 221,137 nodes, 
353,548 edges, and 376,978 retweets. Let $A$ denote the weighted adjacency matrix for $G$. In all cases, weights represent multi-edges, where the multi-edge from node $j$ to node $i$ corresponds to the number of retweets by account $j$ of any tweet by account $i$ in our data set.


\subsection{Degree distributions}
\label{sect:degree}

\begin{figure}[ht!]
    \begin{subfigure}[b]{0.5\textwidth}
        \includegraphics[width=\textwidth]{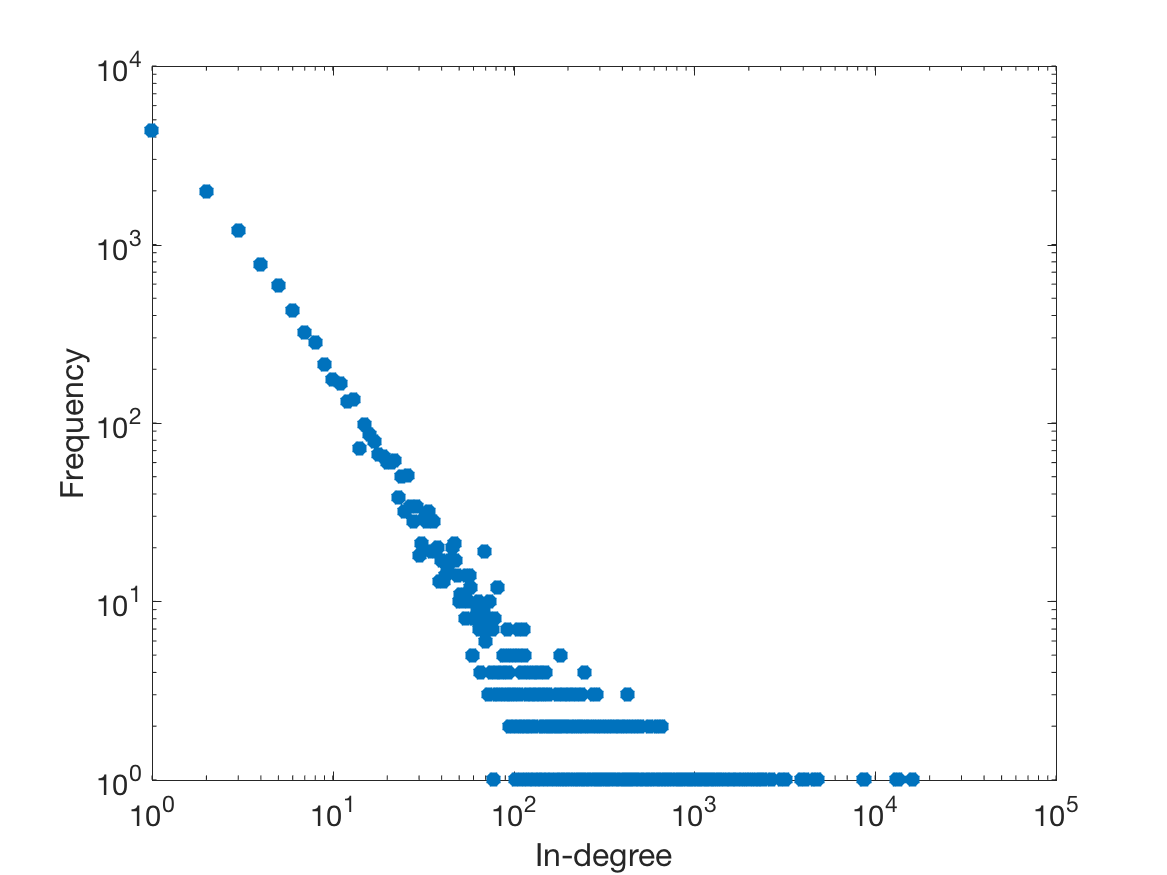}
        \caption{In-degree distribution}
        \label{fig:indeg}
    \end{subfigure}%
    \begin{subfigure}[b]{0.5\textwidth}
        \includegraphics[width=\textwidth]{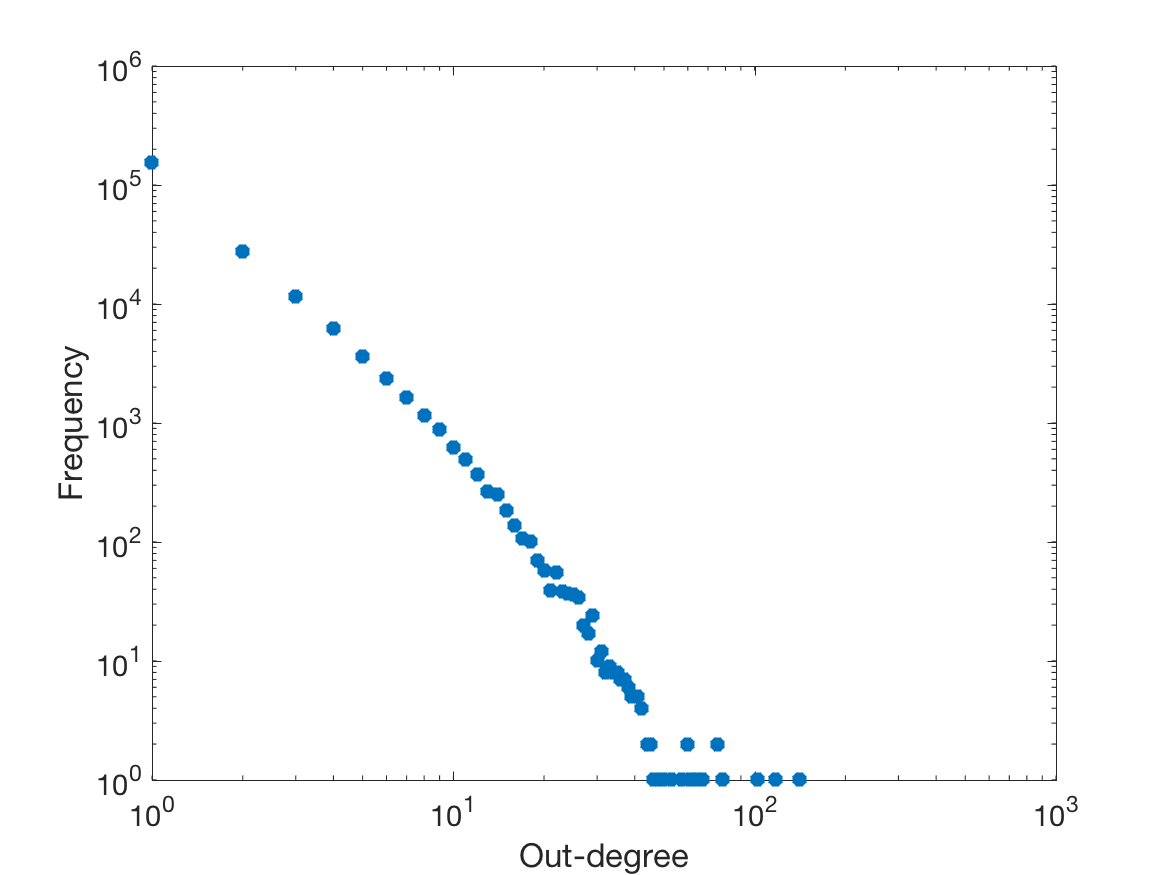}
        \caption{Out-degree distribution}
        \label{fig:outdeg}
    \end{subfigure}
    \caption{Degree distributions for the retweet network $G$.  In-degree represents the number of times that a node was retweeted, and out-degree represents the number of times that a node posted a retweet. The two distributions differ from each other, with the in-degree distribution having a longer tail (corresponding to a few accounts that were retweeted very heavily).}
    \label{fig:degree}
\end{figure}

\begin{figure}[ht!]
    \begin{subfigure}[b]{0.5\textwidth}
        \includegraphics[width=\textwidth]{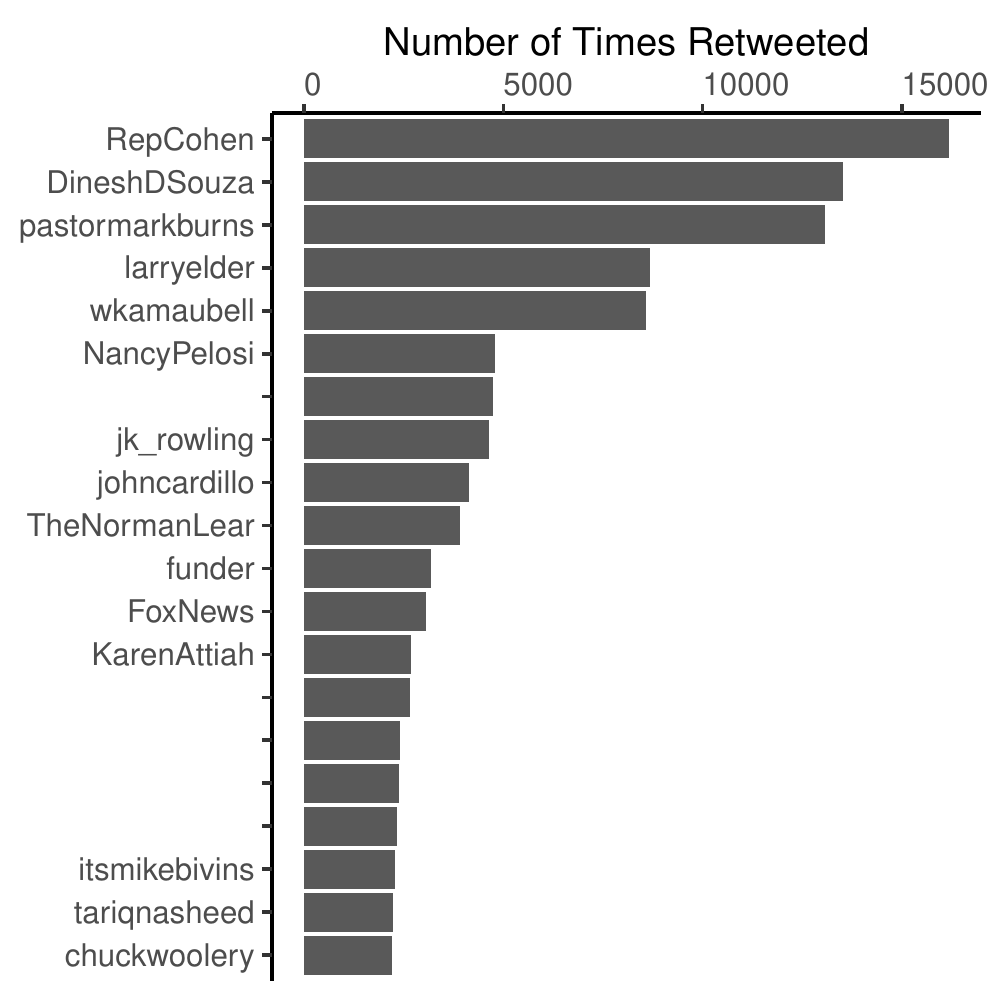}
        \caption{Most retweeted accounts}
        \label{fig:indeg_top}
    \end{subfigure}%
    \begin{subfigure}[b]{0.5\textwidth}
        \includegraphics[width=\textwidth]{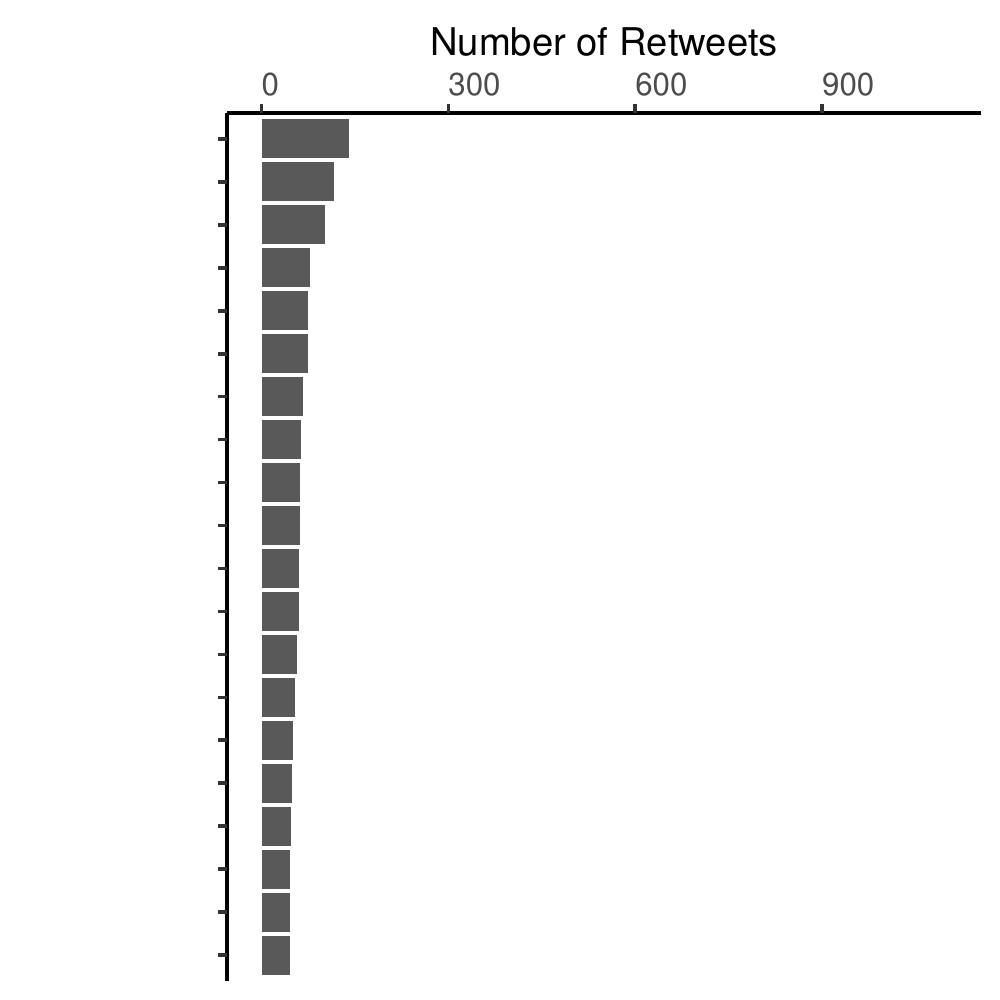}
        \caption{Accounts with the most retweets}
        \label{fig:outdeg_top}
    \end{subfigure}
    \caption{The $20$ nodes that (a) were retweeted the most (i.e., with the largest in-degrees) and (b) posted the most retweets (i.e., with the largest out-degrees). We also show the corresponding in-degrees and out-degrees, respectively. The largest in-degrees are much larger than the largest out-degrees, although the vast majority (94\%) of nodes were never retweeted at all. We show the account names (i.e., handles) for verified accounts; blank labels correspond to accounts that are not verified. The majority of accounts in panel (a) are verified accounts, whereas none of the nodes that posted the most retweets (i.e., the accounts in panel (b)) are verified accounts.
     }      
    \label{fig:degree_top}
\end{figure}

The out-degree of node $k$ corresponds to the total number of retweets that were posted by node $k$, and the in-degree of $k$ corresponds to the total number of times that node $k$ was retweeted.  Unless we specifically note otherwise, we include weights when calculating the in-degrees and out-degrees (i.e., we count all edges in a multi-edge). For example, $\sum_{i=1}^n A_{ij}$ gives the out-degree of node $j$, and $\sum_{j=1}^n A_{ij}$ gives the in-degree of node $i$. In Figure \ref{fig:degree}, we show the in-degree and out-degree distributions for $G$. 
The two distributions differ markedly, as the in-degree distribution has a much longer tail (corresponding to a few accounts that were retweeted very heavily).

In Figure \ref{fig:degree_top}a, we show the in-degrees for the twenty most heavily retweeted accounts.
The mean in-degree is $1.70$, and the standard deviation is $69.22$, indicating extreme heterogeneity in the number of times retweeted. The account (\ttt{RepCohen}) with the largest in-degree was retweeted 16,180 times. By contrast, 208,241 nodes (i.e., 94\% of them) in $G$ were never retweeted at all. We also observe heterogeneity in the out-degree, but it is much less extreme than for in-degree; the standard deviation is $4.89$.  (By definition, the mean in-degree and mean out-degree are the same, as every edge has both an origin and terminus in $G$.) The account with the largest out-degree sent $141$ retweets in our data set.  By contrast, 7,852 accounts had an out-degree of $0$; these accounts were retweeted, but they did not retweet any accounts. In Figure \ref{fig:degree_top}b, we show the twenty accounts that sent the most retweets.

We also consider the in-degree and out-degree distributions for accounts with and without media PCA scores to examine whether there are systematic differences between the two types of accounts. (The former are the 99,412 accounts that followed at least one of the 13 focal media sources.) The heterogeneity in the in-degree distribution that we observed when examining all nodes in $G$ is also present when we consider the in-degree distribution separately for nodes with and without media PCA scores; the standard deviation is $105.24$ for nodes with media PCA and $36.65$ for nodes without it. The mean in-degree for nodes with a media PCA score is larger than for nodes without one ($2.85$ versus $1.08$). Nodes with large in-degree with media PCA scores include \ttt{DineshDSouza}, \ttt{pastormarkburns}, \ttt{RepCohen}, \ttt{wkamaubell}, and \ttt{johncardillo}. However, there are also some heavily retweeted nodes --- such as \ttt{larryelder}, \ttt{TheNormanLear}, and \ttt{NancyPelosi} --- that do not follow any of the $13$ media accounts that we used for computing media PCA scores. We thus cannot compute media PCA scores for these nodes.


\subsection{Centralities}
\label{sect:centrality}

We now examine important accounts by computing several centrality measures \cite{newman2010,baek2019}. We start with degree (i.e., degree centrality), the simplest way of trying to measure a node's importance. In Figure \ref{fig:degree_top}, we show the twenty nodes with the largest in-degrees and the twenty nodes with the largest out-degrees. These two sets are disjoint, indicating that the nodes that generated most of the original content in the Twitter conversation about \#Charlottesville were distinct from those that were most active in promoting existing content through retweets. Degree is a local centrality measure that does not take into account any characteristics of neighboring nodes. For comparison, we also calculate two other widely-used centrality measures, PageRank \cite{brin1998} and HITS \cite{kleinberg1999}, that take some non-local information into account.  

\begin{figure}[ht!]
\centering
\includegraphics[width=0.8\textwidth]{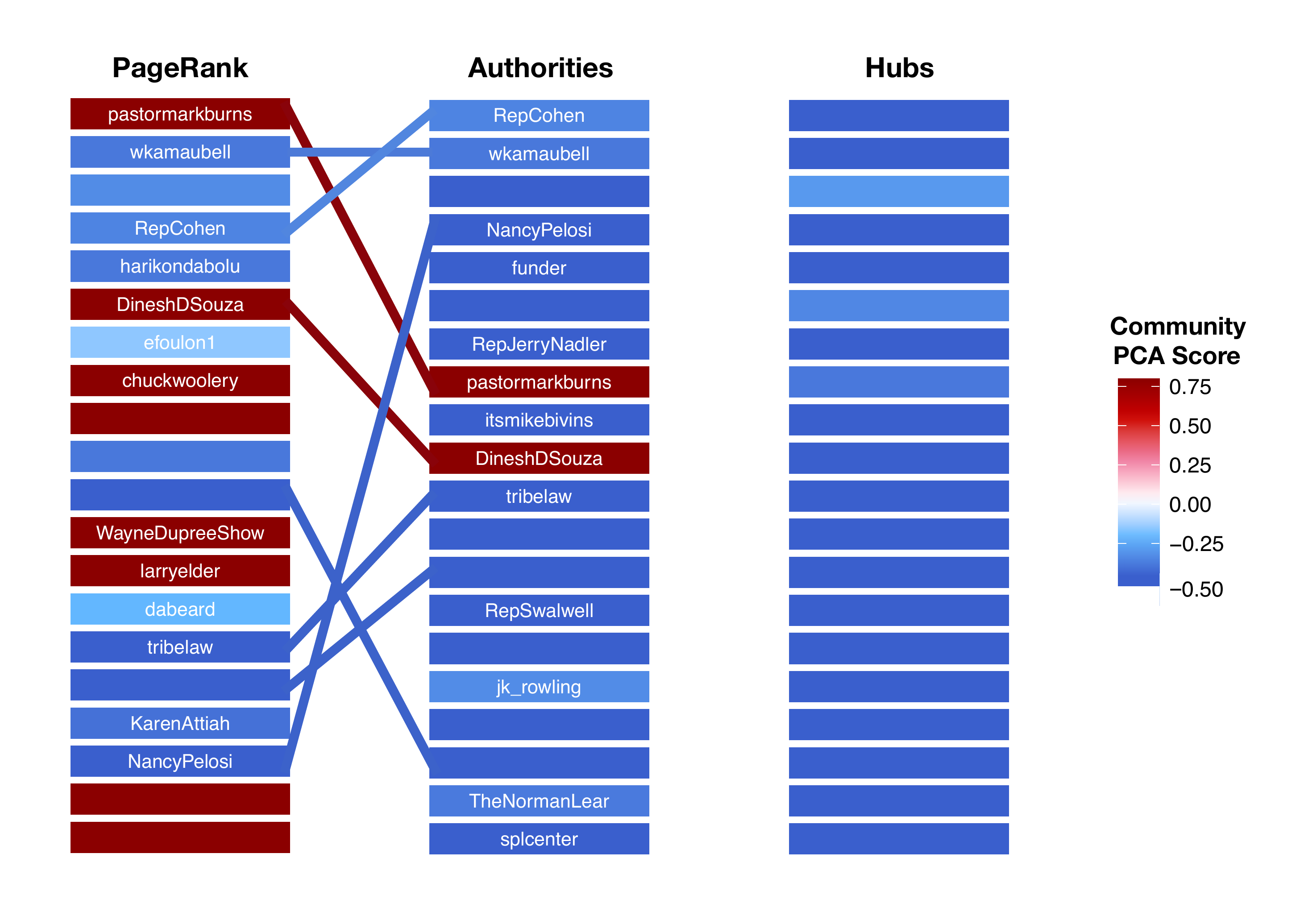}
\caption{The twenty most central nodes in the largest weakly connected component of the retweet network according to (left) PageRank, (center) authorities, and (right) hubs. Color corresponds to the mean media PCA score for the community assignment of each node from modularity maximization using a Louvain method (see Section \ref{sect:communities}). We use colored lines to link nodes that appear in more than one column. There is overlap between the nodes with the largest scores for PageRank and authority centrality, but these two sets are disjoint from the accounts with the largest hub centralities.  All of the best hubs belong to communities with negative (i.e., Left) mean media PCA scores.
}
\label{fig:centrality}
\end{figure}

One obtains PageRank scores from the stationary distribution of a random walk on a network that combines transitions according to network structure and `teleportation' according to a user-supplied distribution \cite{gleich2015_pagerank}, with a parameter that determines the relative weightings of these two processes. We compute PageRank with standard uniform-at-random teleportation using {\sc Matlab}'s \ttt{centrality} function with the default damping factor of $0.85$ (so teleportation occurs for 15\% of the steps in the associated random walk). In the left column of Figure \ref{fig:centrality}, we list the twenty most central nodes according to PageRank. Nine of the these nodes are also on our list of nodes with the largest in-degrees. An exception is \ttt{harikondobalu}, which was retweeted only $38$ times in our data set. The large PageRank value for \ttt{harikondobalu}, despite its small in-degree, reflects the fact that \ttt{harikondobalu} was one of only two nodes that were retweeted by \ttt{wkamaubell}, which was retweeted 8,582 times.

Hub and authority centralities \cite{kleinberg1999} are another useful pair of centrality measures. Using the HITS algorithm, one can simultaneously examine {\it hubs} and {\it authorities}. As discussed in \cite{kleinberg1999}, a good hub tends to point to good authorities, and a good authority tends to have good hubs that point to it. In the context of retweeting, we expect that accounts with large authority scores tend to be retweeted by accounts with large hub scores, and we expect that good hub accounts tend to retweet accounts that are good authorities. As in PageRank, the importances of adjacent nodes influence a node's hub and authority scores. With our convention that the $(i,j)$ entry of a graph's adjacency matrix corresponds to the edge weight from $j$ to $i$, hub and authority scores correspond, respectively, to the principal right eigenvectors of $A^tA$ and $AA^t$. We compute hubs and authorities using {\sc Matlab}'s \ttt{centrality} function.

We list the twenty nodes with the largest authority and hub scores, respectively, in the center and right columns of Figure \ref{fig:centrality}. Color indicates the mean media PCA scores for the community assignment of each account from modularity maximization using a Louvain method \cite{blondel2008,newman2004,newman2006b} (see Section \ref{sect:communities}).
Only two of the nodes among the top-twenty authorities are in communities with positive (i.e., Right) media PCA scores; these accounts, \ttt{pastormarkburns} and \ttt{DineshDSouza}, belong to two prominent conservative personalities.
Neither \ttt{pastormarkburns} nor \ttt{DineshDSouza} were ever retweeted by any of the top-$50$ hubs. By contrast, all of the other authorities were retweeted at least three times by the leading hubs. When we consider all nodes, we observe that the hub scores have a bimodal distribution, with a clear separation between the nodes with small and large values (e.g., using $4\times 10^{-5}$ as a threshold hub score to separate `small' and `large' values). 
We refer to nodes with hub scores that are larger than $4 \times 10^{-5}$ as `large hub-score nodes'. Consider the set of nodes that retweeted \ttt{DineshDSouza}. Of these, the fraction that are large hub-score nodes is $9.0 \times 10^{-4}$. The fraction of nodes that retweeted \ttt{pastormarkburns} that are large hub-score nodes is $1.0 \times 10^{-3}$. The fraction of nodes that retweeted \ttt{itsmikebivins} that are large hub-score nodes is $1.3 \times 10^{-2}$. A few other examples of such fractions are $0.05$ for \ttt{wkamaubell}, $0.15$ for \ttt{tribelaw}, and $1$ for \ttt{RepCohen}.

As is standard for hub and authority scores, there are two qualitatively different ways for a node to have a large authority score: it can either be retweeted many times (e.g., \ttt{DineshDSouza}), or it can be retweeted by nodes with large hub scores (e.g., \ttt{itsmikebivins}). Both of the large-authority Right accounts (\ttt{DineshDSouza} and \ttt{pastormarkburns}) lie in the former category.   

Figure \ref{fig:centrality} also allows us to compare important accounts according to different centrality measures.  As one can see in Figure \ref{fig:centrality}, there is some overlap between the top-PageRank and top-authority accounts.  Note, however, that fewer than half of the top-PageRank accounts are also among the top-authority accounts. By comparison, the set of top hubs is disjoint from the top-PageRank and top-authority accounts in Figure \ref{fig:centrality}.  Additionally, more than half of the top-PageRank and top-authority accounts in Figure \ref{fig:centrality} are verified accounts, whereas none of the top-hub accounts are verified accounts.

%

\subsection{Community structure} \label{sect:communities}

To examine large-scale structure in the \#Charlottesville retweet network, we use community detection to identify tightly-knit sets (so-called `communities') of accounts with relatively sparse connections between these sets \cite{porter2009,fortunato2016}. In our investigation, we employ two widely-used community-detection methods: modularity maximization \cite{newman2004,newman2006b} and InfoMap \cite{rosvall2008}. A major challenge in community detection is parsing what results reflect a network's features, rather than artifacts from a community-detection method.  Modularity maximization and InfoMap are two methods, which use rather different approaches from each other, that have been used successfully on a wide variety of problems. We expect that broad structural features in a network that we observe using both of these methods are likely to be robust to the particular choice of community-detection method, so we expect them to be actual features of the data (rather than artifacts). There exist many other community-detection methods, including statistical inference via stochastic block models \cite{peixoto2017bayesian} and local methods based on personalized PageRank \cite{jeub2015,gleich2015_pagerank}. Exploring our retweet network with other community-detection methods is outside the scope of the present article, but we encourage readers to explore our data set with them. It is available at \url{https://osf.io/487fw/}.


\subsubsection{Modularity maximization}

The modularity of a particular assignment of a network's nodes into communities measures the amount of intra-community edge weight, relative to what one would expect at random under some null model \cite{newman2004,newman2006b}.
Modularity maximization treats community detection as an optimization problem by seeking an assignment of nodes into communities that maximizes a modularity objective function. 
A version of modularity for weighted, directed graphs is \cite{arenas2007,leicht2008}
\begin{equation} \label{eqn:Q}
	Q = \frac{1}{w} \sum_{i=1}^n \sum_{j=1}^n \left( A_{ij} - \gamma \frac{w_i^{\mathrm{in}}w_j^{\mathrm{out}}}{w} \right) \delta (C_i,C_j)\,,
\end{equation}
where
\begin{equation}
	w = \sum_{i=1}^n \sum_{j=1}^n A_{ij}
\end{equation}
is the sum of all edge weights in a network; $w_k^{\mathrm{in}}$ and $w_k^{\mathrm{out}}$ are the in-strength (i.e., a weighted generalization of in-degree) and out-strength (i.e., weighted out-degree), respectively, of node $k$; the community assignment of node $k$ is $C_k$; the quantity $\delta$ is the Kronecker delta; and  $\gamma$ is a resolution parameter that controls the relative weight given to the null model \cite{Lambiotte2009}. Our null-model matrix elements are $P_{ij} = \frac{w_i^{\mathrm{in}}w_j^{\mathrm{out}}}{w}$, so this null model is a type of configuration model \cite{fosdick2018}, in which we preserve expected in-strength and expected out-strength but otherwise randomize connections \cite{porter2009}. For most of our computations, we use the resolution-parameter value $\gamma = 1$ as a default. However, in Section \ref{sect:gamma}, we compare results using a variety of values of $\gamma$ spanning three orders of magnitude. 

To maximize $Q$, we use a {\sc GenLouvain} variant \cite{genlouvain} (which is implemented in {\sc Matlab} and was released originally in conjunction with \cite{mucha2010}) of the locally-greedy Louvain algorithm \cite{blondel2008}. To use the code from \cite{genlouvain}, we symmetrize the modularity matrix $B$, where $B_{ij} = A_{ij} - \gamma \frac{w_i^{\mathrm{in}} w_j^{\mathrm{out}}}{w}$. 
As discussed in \cite{leicht2008}, this is distinct from symmetrizing the adjacency matrix $A$. 

Modularity maximization using {\sc GenLouvain} yields $228$ communities, which range in size from $2$ nodes to 47,321 nodes.
 

\subsubsection{InfoMap}

InfoMap is a community-detection method that is based on the flow of random walkers on graphs \cite{rosvall2008}.\footnote{One can also interpret modularity maximization in terms of random walks on graphs \cite{Lambiotte2009}.} It uses the intuition that a random walker tends to be trapped for long periods of time within tightly-knit sets of nodes \cite{fortunato2016}. Rosvall and Bergstrom \cite{rosvall2008} made this idea concrete by trying to minimize the expected description length of a random walk. For example, one can obtain a concise description of a random walk by allowing node names to be reused between communities. One can apply InfoMap to weighted, directed graphs; and it has been used previously to study Twitter data \cite{weng2013}. To study a directed graph, one introduces a teleportation parameter (as in PageRank); we use the default teleportation value of $\tau=0.15$ \cite{rosvall2008}.

Our implementation uses code from \cite{infomap_web}. With InfoMap, we find $205$ communities, which range in size from $1$ node to 122,504 nodes.


\subsubsection{Large-scale structure of the retweet network}

Several features are evident in our community-detection results from both modularity maximization and InfoMap: (1) communities are largely segregated by media PCA score; (2) overall, the communities skew towards the Left; and (3) most of the nodes on the Right are assigned to a large community that includes prominent right-wing personalities and \ttt{FoxNews}. 

\begin{figure}[h]
\centering
    \begin{subfigure}[b]{0.5\textwidth}
        \includegraphics[width=\textwidth]{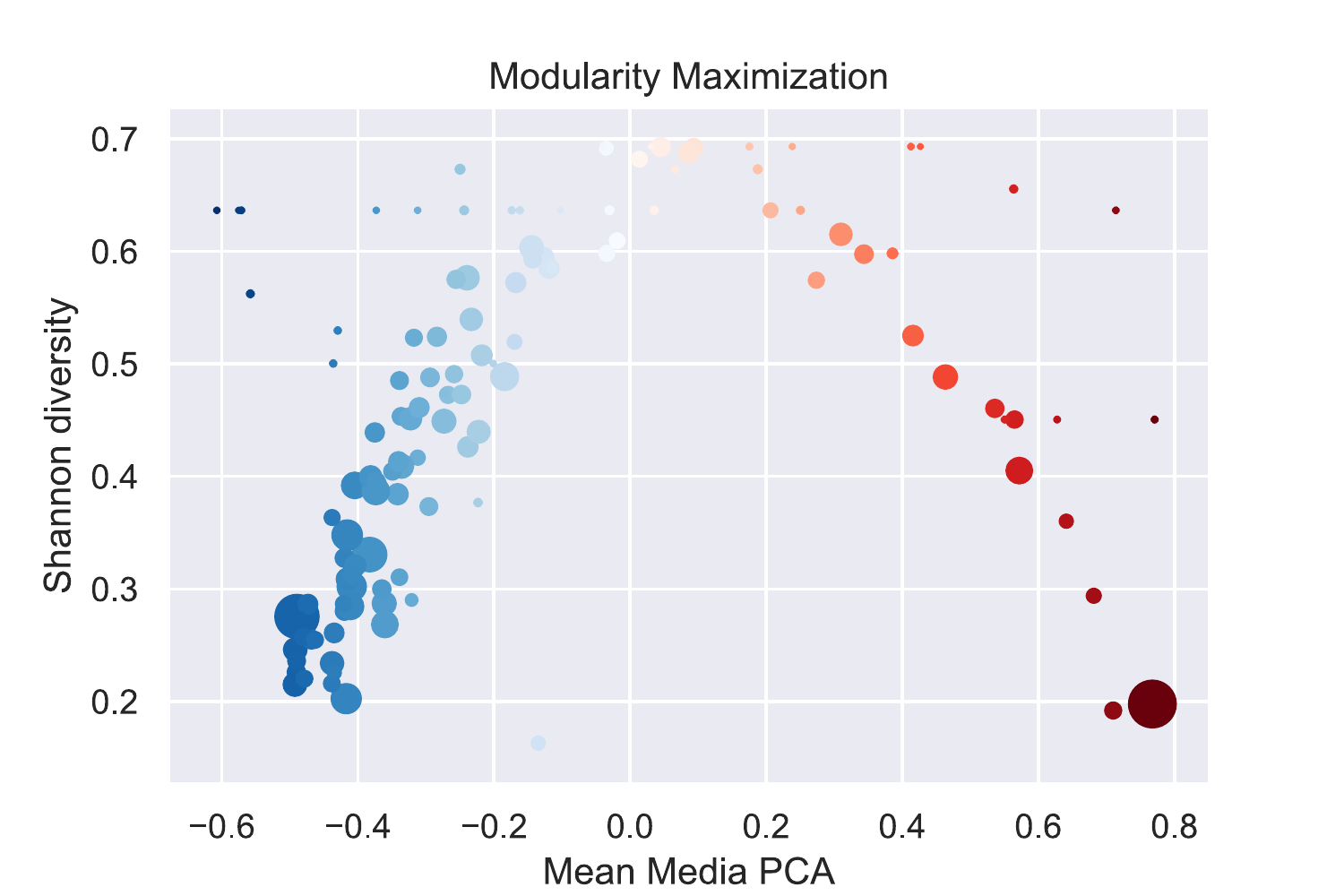}
        \caption{}
        \label{fig:shannon-mod}
    \end{subfigure}%
    \begin{subfigure}[b]{0.5\textwidth}
        \includegraphics[width=\textwidth]{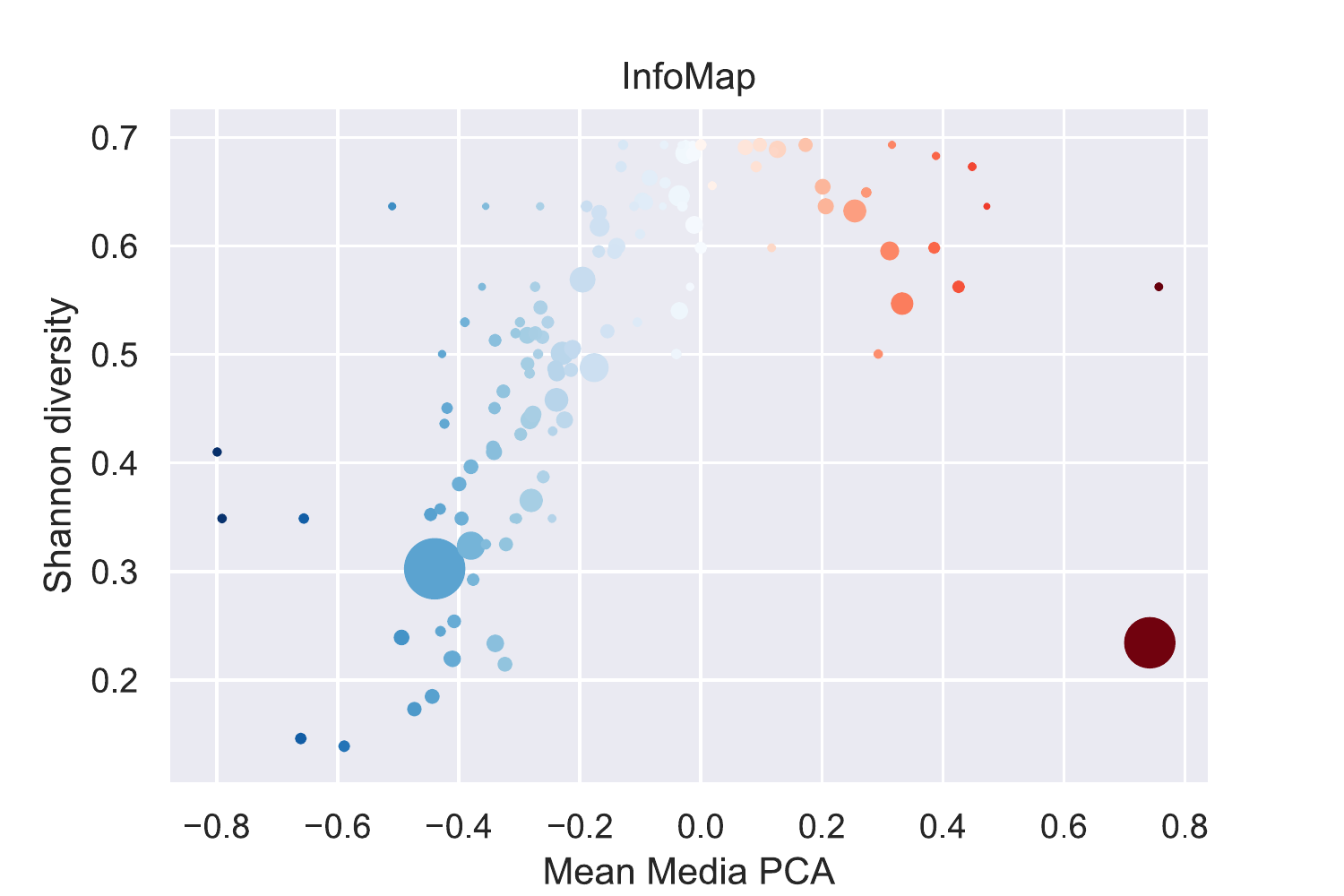}
        \caption{}
        \label{fig:shannon-info}
    \end{subfigure}
    \caption{Shannon diversity index for communities from (a) modularity maximization and (b) InfoMap. The disk sizes correspond to the number of nodes in the communities. More extreme mean media PCA scores are associated with lower diversity in a community, and the larger communities tend to have smaller Shannon Diversities and more polarized media scores. 
    }  
    \label{fig:shannon}
\end{figure}

To examine the relationship between community structure and Left/Right media preference, we compute the mean media PCA score within each community.  The proportion of communities with at least one node with a media PCA score is very similar for modularity maximization (204/228; 89\%) and InfoMap (183/205; 89\%). We also examine the extent of overlap of Left and Right accounts within communities by computing the Shannon diversity index \cite{shannon1948} for each community. This index
\begin{equation}\label{eqn:shannon}
	H^k = -\sum_{i=1}^2 p_i^k \ln p_i^k\,,
\end{equation}
where $H^k$ is the Shannon diversity index for community $k$, and $p_1^k$ and $p_2^k$ (with $p_1^k + p_2^k = 1$ for each $k$) are the fractions of accounts in community $k$ with Left and Right media preferences, respectively. In Figure \ref{fig:shannon}, we show the Shannon diversity scores versus mean media PCA scores for the communities that we detect using modularity maximization and InfoMap.

Both community-detection methods yield a predominantly unimodal shape for the relation between PCA score diversity and mean media PCA score, with more extreme mean media PCA scores associated with lower diversity within a community. Communities with `centrist' mean media PCA scores (i.e., ones that are near $0$) have relatively small sizes. By contrast, the largest communities tend to have mean media scores that are farther from $0$; they also have small Shannon diversity. For example, InfoMap gives two communities that are much larger than the others. One is on the Left (with 122,504 nodes and a mean media PCA score of $-0.43$), and the other is on the Right (with 58,185 nodes and a mean media PCA score of $0.74$). In these two communities, 91\% of the nodes in the largest Left community have negative media PCA scores, compared with 6\% in the largest Right community.  Similarly, the large communities that we obtain from modularity maximization also have little Left/Right node diversity within communities.

Another prominent feature is that both community-detection approaches yield one community on the Right that is much larger than other communities that have a positive mean media PCA score. Additionally, both methods yield a similar set of large-degree accounts in the largest Right community. Specifically, the five nodes with largest in-degrees and out-degrees are the same, with \ttt{DineshDSouza}, \ttt{pastormarkburns}, \ttt{larryelder}, \ttt{johncardillo}, and \ttt{FoxNews} as the five most heavily retweeted accounts (i.e., the ones with the largest in-degrees) in this community.

Our results in Figure \ref{fig:shannon} also suggest that there are more Left-leaning communities than Right-leaning ones.  For example, 106/130 (i.e., about 82\%) of the InfoMap communities with at least ten nodes have negative mean media PCA scores.  Modularity maximization gives a bimodal distribution of community sizes. We refer to communities with at most 100 nodes as `small', communities with more than 100 and at most 1000 nodes as `medium-sized', and communities with more than 1000 nodes as `large'. Of the medium and large modularity-maximization communities, 76/93 (i.e., 82\%) have negative mean media PCA scores. To give some context, we have PCA scores for 78,339 nodes, and 44,797 of them (about 57\%) have a negative media PCA score.


\subsubsection{Finer features of the retweet network}
\label{sect:fine-res}

One notable difference between the two methods is that two large communities dominate for InfoMap (one each on the Left and Right), whereas modularity maximization yields a partition of the retweet network that includes many more large communities. We now examine some of the finer details in the large communities that we obtain from modularity maximization.

Modularity maximization yields $41$ communities with at least 1,001 nodes. To further characterize these $41$ communities, we examine the accounts with the largest in-degrees (i.e., the ones that are retweeted the most) within each community and characterize these nodes by hand from their profiles and, when available (e.g., when account owners are known public personalities), information about the owners of these accounts. More than 85\% (specifically, $35$ of $41$) of these communities have negative (i.e., Left-leaning) mean media PCA scores. The accounts with the largest in-degrees in these $35$ communities include activists (e.g., \ttt{Everytown}, \ttt{IndivisibleTeam}, \ttt{UNHumanRights}, and \ttt{womensmarch}), businesses (e.g., \ttt{benandjerrys}), people from arts and entertainment (e.g., \ttt{jk\_rowling}, \ttt{LatuffCartoons}, \ttt{FallonTonight}, \ttt{ladygaga}, \ttt{Sethrogen}, \ttt{TheNormanLear}, and \ttt{wkamaubell}), journalists (e.g., \ttt{AmyKNelson}), media organizations (e.g., \ttt{AJEnglish}, \ttt{CBSThisMorning}, and \ttt{HuffPostCanada}), and politicians (e.g., \ttt{NancyPelosi}, \ttt{RepCohen}, and \ttt{JoeBiden}). By contrast, only six of the largest communities have positive (i.e., Right-leaning) mean media PCA scores. The largest of these (with 47,321 nodes) includes opinion leaders on the Right (e.g., \ttt{DineshDSouza}, \ttt{pastormarkburns}, and \ttt{larryelder}) and \ttt{FoxNews}, as we discussed previously. Another community has a mean PCA score close to $0$ (specifically, it is $0.086$), and it appears to be a business-oriented community with tweets that are critical of Donald Trump. Two of the remaining four communities with positive media scores are Right-oriented activist communities.  One activist community has $3,987$ nodes, and one of its accounts with among the largest in-degrees (i.e., that is retweeted very heavily) references an influential alt-right account \cite{huffpo-ricky}. The other activist community has 2,710 nodes, and one of its most retweeted accounts references a well-known white supremacist hate symbol in its handle. A third community appears to be a media community with foreign media personalities (e.g., \ttt{KTHopkins}), and the final community of these four is a community that is dominated by accounts that tweet in German. 


\subsubsection{Community characteristics across different resolution-parameter values}
\label{sect:gamma}

To examine the robustness of our findings about community structure in the retweet network, we also conduct modularity maximization using the {\sc GenLouvain} code for a range of values of the resolution parameter $\gamma$ in \eqref{eqn:Q}. There is a `resolution limit' for the smallest detectable community size when using modularity maximization, and the size scales of communities that result from modularity maximization can also influence the sizes of the largest communities \cite{good2010,fortunato2007}. Biases in the sizes of detected communities can skew interpretation of the results of community detection, and distinguishing which results reflect features of a network and which arise from a specific approach or algorithm for community detection is a major challenge. We partially address these concerns by identifying features that are robust to the choice of resolution-parameter value $\gamma$.

\begin{figure}[ht!]
\centering
\includegraphics[width=0.5\textwidth]{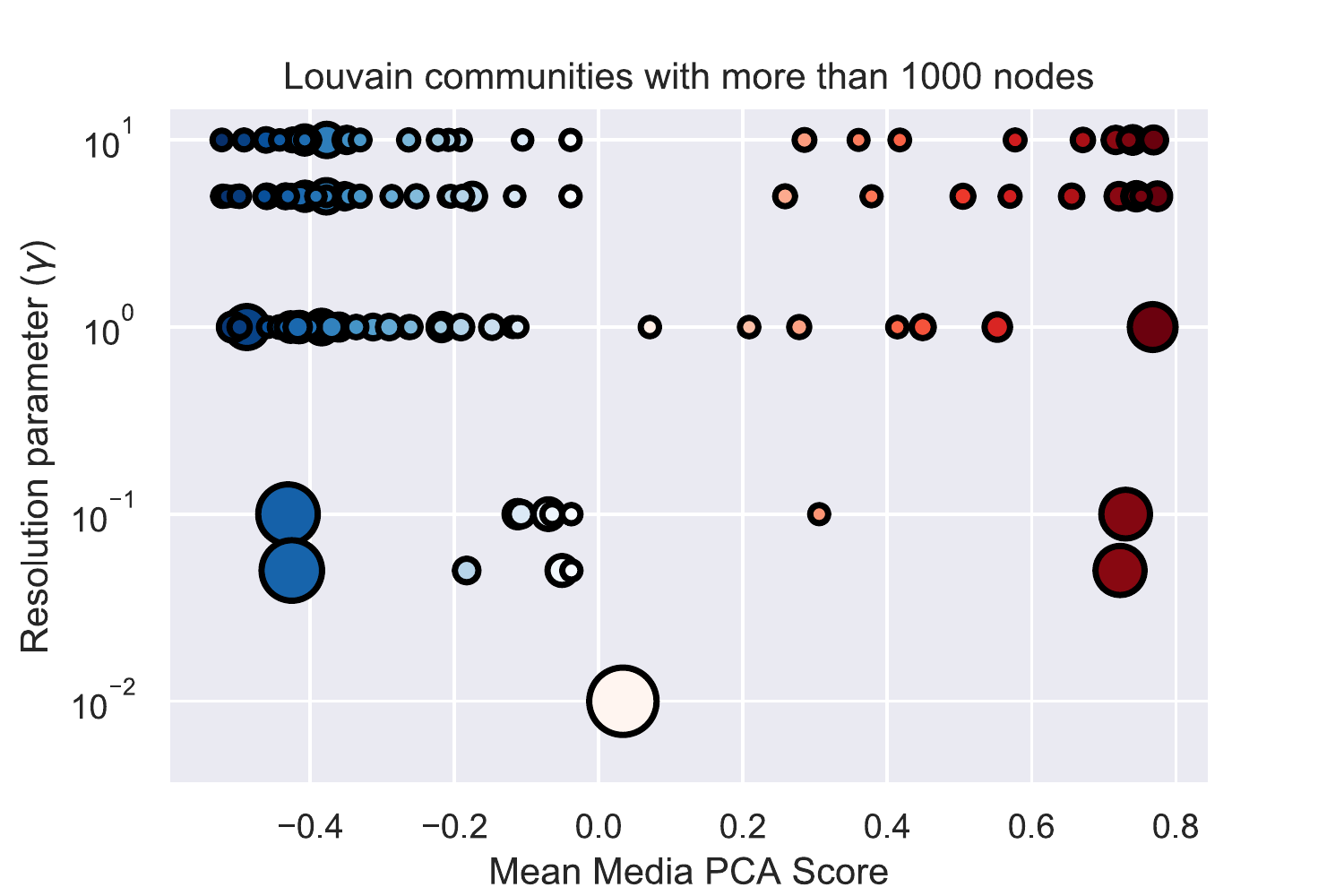}
\caption{Modularity maximization results using the {\sc GenLouvain} code for a range of values of the resolution parameter $\gamma$ ($0.01, 0.05, 0.1, 1, 5,$ and $10$). For each value of $\gamma$, we plot communities with more than $1000$ nodes versus their mean media PCA score. Colors correspond to the mean media PCA score (with, as usual, `Left' signifying negative values and `Right' signifying positive values), and disk size corresponds to the number of nodes in a community. For all explored values of $\gamma$ that result in more than one community, we detect more large communities (i.e., those with more than 1000 nodes) on the Left than on the Right.
 }
\label{fig:comm_gamma}
\end{figure}

{In Figure \ref{fig:comm_gamma}, we show our results from community detection using the {\sc GenLouvain} code with resolution-parameter values
$\gamma$ that range from $10^{-2}$ to $10$. Smaller values of $\gamma$ in \eqref{eqn:Q} favor fewer communities (with 783 communities using $\gamma=10$, compared with only a single community for $\gamma=10^{-2}$). For each value of $\gamma$, we plot communities with more than 1000 nodes versus their mean media PCA score. We observe that there are more large (i.e., having more than 1000 nodes) Left-leaning than Right-leaning communities for all examined values of $\gamma$ for which we detected more than one community, suggesting that this is a robust feature of our Twitter retweet network.

\subsection{Modular centrality}

Our results in Section \ref{sect:communities} suggest that the Charlottesville retweet network has meaningful modules. Ghalmane et al. \cite{ghalmane2019} examined a way to exploit community structure when identifying influential nodes, and they computed different `modular centrality' measures to describe influence within versus between communities. (See, e.g., \cite{guimera2005,fenn2012} for other work on quantifying centrality values within versus between communities.) In this section, we compute a slightly modified version of the  `modular degree centrality' from \cite{ghalmane2019}. We separately count the number of times (producing an `inter-community in-degree') that a given node is retweeted by nodes that belong to communities other than the one that includes that node and the number of retweets (producing an `intra-community in-degree') by nodes that belong to the same community as that node. We then compute the ratio of the inter-community in-degree to the intra-community in-degree.

\begin{table}[ht]
\tiny
\begin{centering}
\begin{tabular}{lc}
 & Inter-community in-degree   \\
Node & divided by intra-community in-degree \\  \hline
\blue{{\tt RepCohen}} & \blue{0.36725} \\ 
\red{{\tt DineshDSouza}} & \red{0.040400} \\ 
\red{{\tt pastormarkburns}} & \red{0.036085} \\ 
\red{{\tt larryelder}} & \red{0.043745} \\ 
\blue{{\tt wkamaubell}} & \blue{0.42938} \\ 
\blue{{\tt NancyPelosi}} & \blue{0.21463} \\ 
	 & \blue{0.25073} \\ 
\blue{{\tt jk\_rowling}} & \blue{0.22067} \\ 
\red{{\tt johncardillo}} & \red{0.078068} \\ 
\blue{{\tt TheNormanLear}} & \blue{0.31365} \\ 
\blue{{\tt funder}} & \blue{0.17688} \\ 
\red{{\tt FoxNews}} & \red{0.077005} \\ 
\blue{{\tt KarenAttiah}} & \blue{0.32803} \\ 
	 & \blue{0.28350} \\ 
	 & \blue{0.23508} \\ 
	 & \red{0.050066} \\ 
	 & \red{0.043146} \\ 
\blue{{\tt itsmikebivins}} & \blue{0.23335} \\
\blue{{\tt tariqnasheed}} & \blue{0.23772} \\ 
\red{{\tt chuckwoolery}} & \red{0.041193} \\ 
\end{tabular}
\caption{Number of inter-community retweets divided by the number of intra-community retweets for the twenty most heavily retweeted nodes in the retweet network. We obtain communities by maximizing modularity using the {\sc GenLouvain} code with a resolution-parameter value of $\gamma = 1$. Color indicates the mean media PCA score of the {\sc GenLouvain community} for each node. We use blue to signify communities that are on the Left (i.e., ones with a negative mean media PCA score) and red to signify communities that are on the Right (i.e., ones with a positive mean media PCA score).  Nodes with large in-degrees on the Left had larger values of the number of inter-community retweets divided by the number of intra-community retweets than was the case for nodes with large in-degrees on the Right. We label the Twitter accounts only of public profiles.
 }
\label{table:mod_centr}
\end{centering}
\end{table}

In Table \ref{table:mod_centr}, we list the ratio of the inter-community in-degree to intra-community in-degree for the twenty nodes with the largest in-degrees in our retweet network. This ratio tends to be significantly larger for nodes that belong to Left-leaning communities (their range is 0.18--0.43, with a mean of $0.27$) than for nodes that belong to Right-leaning communities (their range is 0.036--0.078, with a mean of $0.051$).


\section{Media-preference assortativity}  \label{sec5}

In this section, we quantify the extent to which retweets occur between nodes with similar media preferences.  We do this by examining assortativity in our retweet network in terms of both the value and the sign of the media PCA score. Large assortativity indicates that the Charlottesville conversation largely splits according to media PCA score. Combined with our results of Section \ref{sec6} (in which we compare tweet content on the Left and the Right), this gives a way to assess the extent of polarization in the Twitter conversation.

To examine homophily in media-preference scores in the Twitter conversation about \#Charlottesville, we measure media-preference assortativity by computing the Pearson correlation coefficient of the media PCA score for nodes in the retweet network. Specifically, we compute the correlation of the media PCA score for dyads (i.e., nodes that are adjacent to each other via an edge) in the retweet network. We ignore edge weights, and we restrict our calculations to dyads for which we have a PCA score for both nodes. There are 93,521 such pairs.  

The  correlation coefficient of the media PCA scores is $\rho \approx 0.67$. For comparison, as a null model, we compute the correlation-coefficient distribution for 100,000 random permutations of the PCA scores of the nodes. Specifically, in each realization, we fix the network and assign the PCA scores uniformly at random to the nodes for which PCA scores were available originally. (For an alternative approach for examining assortativity, see \cite{park2007}.) The resulting distribution for the correlation coefficient $\rho$ appears to be approximately Gaussian, with a mean of $-1.29 \times 10^{-5}$ and a standard deviation of $0.0033$. The z-score for the measured correlation coefficient of $0.67$ is larger than $203$, indicating that the retweet network has a statistically significant media-preference assortativity.

We also compute the assortativity coefficient $r$ that was introduced by Newman \cite{newman2002assort,newman2003b}.  Suppose that there are $g$ types of nodes in the network. Following \cite{newman2003b}, we calculate
\begin{equation}\label{eqn:assort}
	r = \frac{\sum_{\ell=1}^g e_{\ell \ell} - \sum_{\ell=1}^ g a_\ell b_\ell}{1 - \sum_{\ell=1}^g a_\ell b_\ell}\,, 
\end{equation}
where $e_{\ell s}$ is the fraction of the edges in a network that emanate from a node of type $\ell$ and terminate at a node of type $s$, the quantity $a_\ell = \sum_{s=1}^g e_{\ell s}$ is the fraction of the edges that emanate from a node of type $\ell$, and $b_s = \sum_{\ell=1}^g e_{\ell s}$ is the fraction of edges that terminate at a node of type $s$.

%
\begin{table}[ht]
\begin{centering}
\begin{tabular}{c|cc}
 & \multicolumn{2}{c}{\it original} \\ 
{\it retweeter} & {\bf Left} & {\bf Right} \\ 
{\bf Left} & 0.43 & 0.057 \\
{\bf Right} & 0.044 & 0.47 \\
\end{tabular}
\caption{Mixing matrix of the fractions of the number of edges that correspond to edges between the different types of accounts, as characterized by the sign of their media PCA score (i.e., first PCA score). As usual, `Left' indicates a negative media PCA score, and `Right' indicates a positive media PCA score. (No nodes have a media PCA score of exactly $0$.) Accounts tend to mix with (i.e., be adjacent to) accounts with a PCA score of the same sign, as indicated by the larger values on the diagonal of the mixing matrix.
}
\label{table:mixing}
\end{centering}
\end{table}

To calculate \eqref{eqn:assort} for our retweet network, we classify nodes according to the sign of their media PCA score. 
In the largest weakly connected component of the retweet network, we have PCA scores for 78,339 nodes, of which 44,797 (i.e., 57\% of them) have a negative media PCA score. The resulting assortativity coefficient is $r \approx 0.80$.  We show the mixing matrix $e$ in Table \ref{table:mixing}. As a comparison, Newman \cite{newman2003b} calculated an assortativity coefficient of $0.62$ by ethnicity for the sexual-partner network that was described in \cite{catania1992}.    

Five\footnote{Specifically, these accounts are \ttt{csmonitor}, \ttt{MotherJones}, \ttt{theblaze}, \ttt{FoxNews}, and \ttt{NPR}.} of the media accounts that we used to compute the media PCA score also appear as nodes in the retweet network $G$.  Of these, \ttt{FoxNews} was retweeted $3049$ times, \ttt{NPR} was retweeted $69$ times, \ttt{MotherJones} was retweeted $15$ times, and \ttt{csmonitor} was retweeted $6$ times. Removing these media accounts from $G$ has a negligible effect on the assortativity coefficient $r$.

Although the assortativity by media PCA score in the retweet network is rather large, there are some prominent exceptions. For example, \ttt{RepCurbelo} and \ttt{SenatorTimScott}\footnote{These are the Twitter accounts for Representative Carlos Curbelo (FL, Republican) and Senator Tim Scott (SC, Republican).}, the accounts for two Republican members of Congress, were heavily retweeted in Left-leaning communities that we detected using modularity maximization. However, both \ttt{RepCurbelo} ($0.49$) and \ttt{SenatorTimScott} ($0.12$) have positive (i.e., Right) media PCA scores, consistent with their affiliation with the Republican party. \ttt{RepCurbelo} was the fourth-most retweeted account in a community from modularity maximization with a negative (i.e., Left) mean media PCA score ($-0.32$). \ttt{RepCurbelo}, who spoke out strongly against the events in Charlottesville \cite{curbelo2017}, was retweeted by 22 accounts. We have PCA scores for 9 of these accounts, of which $4$ have media PCA scores on the Left. Similarly, \ttt{SenatorTimScott} was the second-most retweeted account in a Left-leaning community (with a mean media PCA score of $-0.26$) that we obtained from modularity maximization. \ttt{SenatorTimScott} was retweeted by $78$ accounts, and nearly half (specifically, $20$ of $43$) of the accounts that retweeted \ttt{SenatorTimScott} for which we have PCA scores have negative media PCA scores. We identified \ttt{RepCurbelo} and \ttt{SenatorTimScott} as accounts that warrant examination by first compiling the list of nodes that were retweeted by accounts with media PCA scores of the opposite sign and then examining this list for prominent accounts. One can further develop this approach (for example, to identify negative or mocking retweets \cite{tufekci2014}), and it may be useful in other situations for identifying accounts that generate communication across ideological or other divides.


\section{Comparison of tweet content between Left and Right}  \label{sec6}

In this section, we examine tweet content from Left-leaning and Right-leaning communities. Comparing word and hashtag frequency allows us to see some of the ways in which the Twitter conversation about Charlottesville differed between the Left and the Right.

We use the Python library {\sc nltk 3.3} to tokenize tweets into words and punctuation. In Table \ref{table:wordcounts}, we show the twenty-five most numerous words in our data set. We separately consider accounts with negative (i.e., Left) and positive (i.e., Right) media PCA scores after removing stop words.\footnote{We use stop words from the {\sc nltk} Python library, and we also remove the following words when calculating word counts: `t', `https', `co', `RT', `s', `amp', `n', `w', and `c'.}  We do not stem the words in our data set, and we treat different capitalizations as different words. We find some overlap between the Left and Right data sets. For example, tweets related to `Trump' were very common regardless of media PCA score. `Barcelona' was also one of the most numerous words in tweets by both the Left and the Right. (There was a 17 August 2017 van attack in that city that killed $13$ individuals (at the time of data collection) and injured more than $100$ others.\footnote{One of those wounded individuals died later (after the time of data collection) from their injuries.})  
However, as we can see from the words in Table \ref{table:wordcounts}, there are also many differences in the words that were used by the Left and the Right. We illustrate such differences by coloring the relevant words. For example, `Obama' was the third-most numerous word in tweets that were sent by nodes with positive media PCA scores, but it was not in the top one hundred for nodes with negative media PCA scores. Additionally, `Nazi' appeared commonly in tweets from the Left, but it did not appear often in tweets from the Right. By contrast, the words `Antifa' and `MSM' were used often by the Right but not by the Left.

\begin{table}[ht]
\begin{centering}
\tiny
\begin{tabular}{lclc}
\multicolumn{2}{c}{{\bf Left}} &  \multicolumn{2}{c}{{\bf Right}} \\
Word & Count & Word & Count \\ \hline 
Charlottesville &  98782 & Charlottesville &  84282\\ 
Trump &  19352 & Trump &  11376\\ 
\blue{realDonaldTrump} &  10289 & \red{Obama} &  8195\\ 
white &  9472 & white &  8174\\ 
\blue{Nazis} &  7743 & \red{DineshDSouza} &  8026\\ 
\blue{Nazi} &  6451 & \red{POTUS} &  7614\\ 
\blue{comments} &  5068 & \red{pastormarkburns} &  7394\\ 
charlottesville &  4759 & Barcelona &  6348\\ 
\blue{good} &  4693 & \red{supremacist} &  6004\\ 
\blue{people} &  4637 & \red{organizer} &  5864\\ 
\blue{response} &  4091 & rally &  5851\\ 
\blue{must} &  4080 & violence &  5671\\ 
hate &  3933 & \red{guy} &  5501\\ 
Barcelona &  3930 & \red{MSM} &  5070\\ 
violence &  3884 & hate &  4882\\ 
\blue{supremacy} &  3642 & Right &  4531\\ 
\blue{introducing} &  3584 & \red{city} &  4490\\ 
\blue{attack} &  3460 & \red{Americans} &  4489\\ 
\blue{via} &  3382 & \red{larryelder} &  4421\\ 
\blue{RepCohen} &  3358 & \red{Antifa} &  4409\\ 
rally &  3345 & \red{Since} &  4183\\ 
\blue{Impeachment} &  3341 & \red{11} &  4011\\ 
\blue{Articles} &  3326 & \red{Chicago} &  3946\\ 
\blue{Klansmen} &  3310 & \red{Statues} &  3905\\ 
right &  3155 & \red{40} &  3900\\ 
\end{tabular}
\caption{The twenty-five most numerous words for nodes with negative (i.e., Left) and positive (i.e., Right) media PCA scores. The blue text indicates words that appear in the top-twenty words for the Left but not for the Right, and the red text indicates words that appear in the top twenty for the Right but not for the Left.
}
\label{table:wordcounts}
\end{centering}
\end{table}

\begin{table}[ht]
    \begin{minipage}{.5\linewidth}
      \centering
        \tiny
        \begin{tabular}{lclc}
        \multicolumn{2}{c}{{\bf Left}} &  \multicolumn{2}{c}{{\bf Right}} \\
Word & Count & Word & Count\\\hline
Charlottesville &  30995 & Charlottesville & 16218\\
Trump &  19396 & Trump & 11376\\
realDonaldTrump &  10291 & realDonaldTrump & 3245\\
\blue{Nazis} &  5600 & \red{MAGA} & 1403\\
comments &  4941 & \red{POTUS} & 1278\\
\blue{good} &  3788 & \red{President} & 1229\\
\blue{introducing} &  3580 & \red{Romney} & 1209\\
\blue{Impeachment} &  3331 & comments & 1175\\
\blue{white} &  3318 & \red{apologize} & 1164\\
\blue{Articles} &  3313 & \red{racist} & 1149\\
\blue{RepCohen} &  3306 & \red{blame} & 1096\\
\blue{Klansmen} &  3301 & \red{Mayor} & 1085\\
\blue{must} &  2699 & \red{antifa} & 980\\
\blue{Congress} &  2606 & charlottesville & 959\\
\blue{censure} &  2540 & \red{Vice} & 958\\
\blue{supremacy} &  2239 & \red{Barcelona} & 880\\
\blue{wake} &  2107 & \red{left} & 868\\
\blue{defense} &  2060 & \red{alt} & 818\\
\blue{NancyPelosi} &  2054 & \red{coming} & 781 \\
\blue{repulsive} &  2018 & \red{non} & 776 \\
        \end{tabular}
    \end{minipage}%
    \begin{minipage}{.5\linewidth}
      \centering
        \tiny
        \begin{tabular}{lclc}
        \multicolumn{2}{c}{{\bf Left}} &  \multicolumn{2}{c}{{\bf Right}} \\
Word & Count & Word & Count \\ \hline
Charlottesville &  4213 & Charlottesville &  7384\\
Barcelona &  3926 & Barcelona &  6345\\
attack &  1023 & \red{Muslims} &  2101\\
Trump &  1020 & \red{13} &  2036\\
\blue{terrorism} &  842 & \red{right} &  2035\\
\blue{2} &  665 & \red{CNN} &  2019\\
\blue{prayers} &  450 & \red{kills} &  2006\\
\blue{realDonaldTrump} &  438 & \red{condemned} &  2002\\
\blue{thoughts} &  420 & \red{kill} &  1996\\
\blue{Terror} &  415 & \red{someone} &  1977\\
\blue{gets} &  400 & attack &  1707\\
\blue{settle} &  399 & \red{left} &  1686\\
\blue{directed} &  399 & \red{lunatic} &  1632\\
\blue{scolding} &  399 & \red{One} &  1621\\
\blue{intentional} &  397 & \red{wholesale} &  1620\\
\blue{ambigu} &  395 & \red{johncardillo} &  1617\\
\blue{condemns} &  356 & \red{copycat} &  1190\\
\blue{took} &  346 & Trump &  718\\
\blue{comment} & 317 & \red{Blitzer} &  648\\
\blue{immediately} & 317 & \red{people} &  642\\
               \end{tabular}
    \end{minipage} 
        \caption{The twenty most numerous words (excluding words that correspond to handles of non-verified accounts) from the subset of tweets that include (left set of columns) the word `Trump' and (right set of columns) the word `Barcelona'. The blue text indicates words that appear in the top-twenty words for the Left but not for the Right, and the red text indicates words that appear in the top twenty for the Right but not for the Left.
}
    \label{table:wordcounts-sub}
\end{table}

We also observe other qualitative differences between the tweet content of the Left and Right on shared common words, such as `Trump' and `Barcelona', in the \#Charlottesville data set. The `Trump' subset\footnote{Aside from `Charlottesville', which comes directly from the hashtag that we used to generate the data set, the word `Trump' was the most common word for both Left and Right.} for which we have media PCA scores consists of 34,084 total tweets (of which about 32\% are unique) from the Left and 18,791 total tweets (of which about 23\% are unique) from the Right.\footnote{Duplicate tweets arise, for example, from retweets.  It is also possible that multiple accounts independently posted identical content that were not retweets.} As we show in the left set of columns of Table \ref{table:wordcounts-sub}, the Left and Right conversations about Trump differ markedly from each other. The `Barcelona' subset consists of 4779 tweets (of which 1672 are unique) from the Left and 7669 tweets (of which 1401 are unique) from the Right.  
In the right set of columns of Table \ref{table:wordcounts-sub}, we show the twenty most numerous words for the Barcelona subset for both Left and Right. For the Right, our examination of the most heavily retweeted tweets suggests that much of the discussion about `Barcelona' in our data set involves comparing media coverage of the Charlottesville and Barcelona attacks. On the Left, some of the heavily retweeted tweets about `Barcelona' centered on comparing Trump's reaction to the Barcelona and Charlottesville attacks.

We also implement an idea from Gentzkow and Shapiro \cite{gentzkow2010}, who used a chi-square statistic to analyze the different phrase usage of Democrats and Republicans in Congressional speeches. We apply their approach to words in tweets from the Left and Right in the `Trump' subset (specifically, using equation (1) in \cite{gentzkow2010} with `phrases' that consist of a single word) and find that the five words (which include `Nazis', `antifa', and `Vice') with the largest chi-square values were also among the most common words in the `Trump' subset (see Table \ref{table:wordcounts-sub}). Therefore, we observe some consistency in results across different methods.

\begin{figure}[ht!]
\centering
\includegraphics[width=0.8\textwidth]{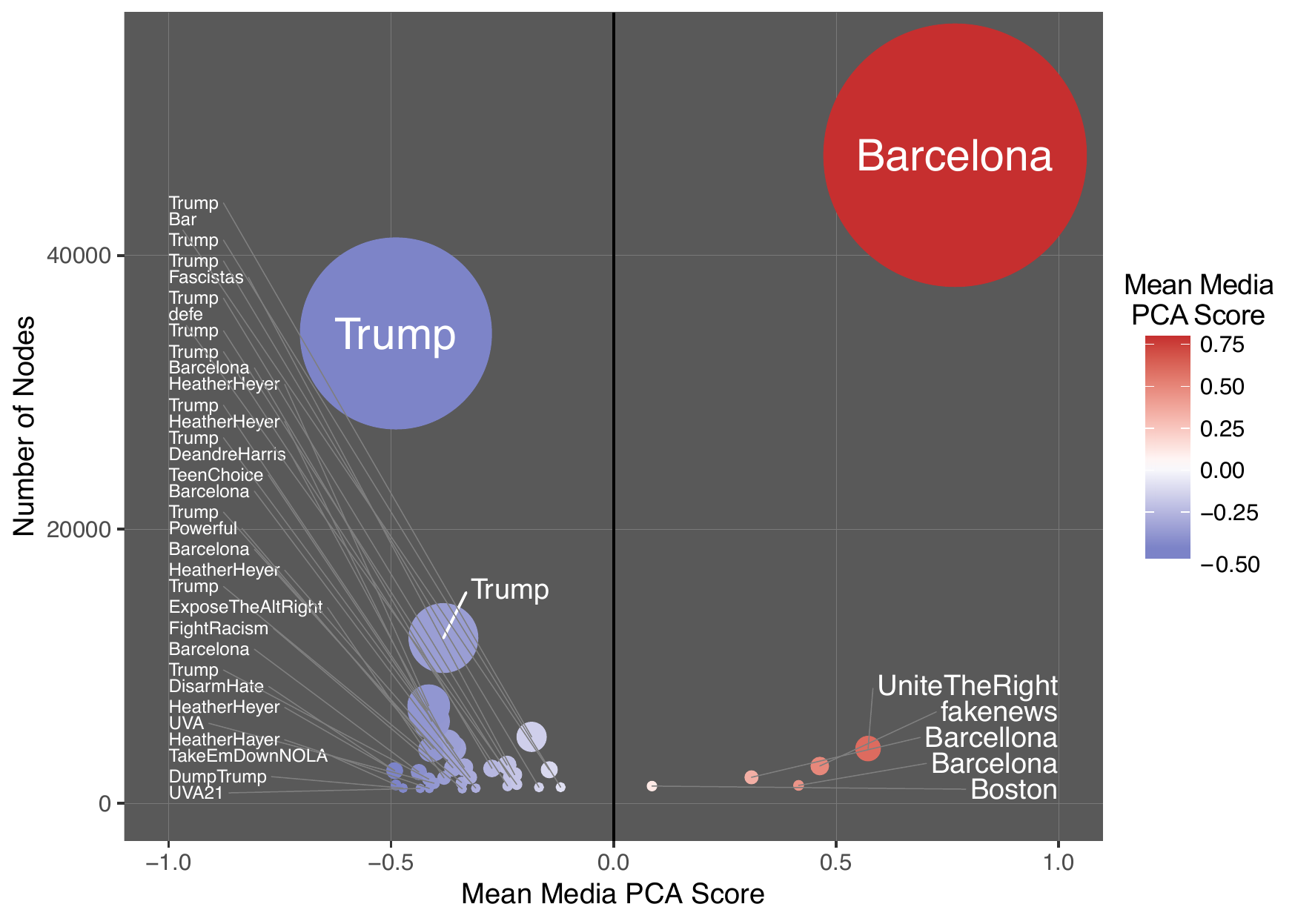}
\caption{Most numerous hashtags for communities from modularity maximization with at least 1,001 nodes. The mean media PCA score is on the horizontal axis (and is also indicated by the color bar), and the number of nodes is on the vertical axis.
}
\label{fig:comm_hashtags}
\end{figure}

We also use hashtags to compare tweets between the Left and Right communities. In Figure~\ref{fig:comm_hashtags}, we show the most numerous hashtag for each community,\footnote{This neglects hashtags that include `Charlottesville', as our data collection was based on the \#Charlottesville hashtag.} together with the community's mean media PCA score. On the Left, the most numerous hashtag is \#Trump (in $13$ of $35$ communities), followed by \#HeatherHeyer (in $5$ of $35$ communities, if we include a single community with `\#HeatherHayer') and then \#Barcelona (in $4$ of $35$ communities). Other top hashtags include \#ExposeTheAltRight, \#DumpTrump, \#FightRacism, and \#DisarmHate.  On the Right, \#Barcelona is the most numerous hashtag (in $3$ of $6$ communities, if we include a single community with \#Barcellona).  Other top hashtags on the Right are \#UniteTheRight (from a community with an account of large in-degree whose Twitter handle references an influential account that identifies with the alt-right \cite{huffpo-ricky}) and \#fakenews (from a community with an account of large in-degree whose handle references a well-known white-supremacist hate symbol).

The fractions of tweets with unique content also differ between the Left and Right. Nodes with negative media PCA scores posted a total of 112,314 tweets, of which 42,458 (i.e., about 38\%) were unique. Nodes with positive media PCA scores posted a total of 92,575 tweets, of which 22,462 (i.e., about 24\%) are unique. We also observe a larger fraction of original content in the Left than in the Right when restricting to several specific topics, including `Trump' (32\% for the Left versus 23\% for the Right), `Barcelona' (35\% for the Left versus 18\% for the Right), `MSM' (21\% for the Left versus 8\% for the Right), `Obama' (31\% for the Left versus 7\% for the Right), and `Antifa' (50\% for the Left versus 22\% for the Right). However, there are a slightly larger proportion of unique tweets for the Right for tweets that include the word `Nazi' (21\% for the Left versus 26\% for the Right).


\section{Conclusions and discussion}  \label{sec7}

Our study of the Twitter conversation about \#Charlottesville illustrates that (1) one can reasonably characterize nodes in terms of their media followership and a one-dimensional PCA-based Left/Right orientation score, (2) the Charlottesville retweet network is highly polarized with respect to this measure of Left/Right orientation, and (3) communities in the retweet network are largely homogeneous in their Left/Right node composition. Our findings thus indicate that, with a few exceptions, the Twitter conversation largely split along ideological lines, as measured by the media preference of Twitter accounts.

As we just summarized, our investigation illustrates strong polarization in the Twitter conversation about \#Charlottesville. 
We found that media followership on Twitter is informative and that the \#Charlottesville retweet network is strongly assortative with respect to a corresponding PCA-based Left/Right orientation score. Our finding of positive assortativity with respect to media preference on Twitter is also consistent with previous studies of Twitter data \cite{colleoni2014,conover2011,feller2011,pennacchiotti2011}. However, in contrast with these previous studies, our approach to node characterization does not require text analysis or labeled data for training.  Characterizing nodes via a principal component analysis of media followership is simple, easy to interpret, and provides a valuable complement to characterizing nodes based on the content of their tweets. Because the \#Charlottesville retweet network is strongly assortative with respect to media preference, it is a potentially useful indicator of marked polarization on Twitter about the `Unite the Right' rally and its aftermath. However, whether differences in media preferences are a cause or an effect (or both) of assortativity on social media is not something that our approach allows us to conclude.

Polarization is also evident in the community structure of the retweet network, as the communities are highly segregated in terms of their Left/Right node composition. The Left has a larger proportion of tweets with original content (as opposed to retweets) than the Right, and nodes with large hub scores tended to retweet nodes on the Left rather than those on the Right. We also found that modularity maximization detects Left communities with central nodes from disparate focal areas (such as business, media, entertainment, and politics). A robust feature of our community-detection results is that there are more large Left-leaning communities than Right-leaning ones.  We also found that Left nodes with high in-degree had larger ratios of inter-community in-degrees to intra-community in-degrees, suggesting that heavily retweeted nodes on the Left were more likely to be retweeted by communities other than their own. Taken together, these findings suggest that Twitter accounts on the Left that condemned the `Unite the Right' rally in Charlottesville and Trump's handling of the aftermath came from broad segments of society.   Support on the Right in our data was concentrated in fewer communities, the largest of which includes the five most heavily retweeted accounts on the Right: \ttt{FoxNews} and right-wing personas  \ttt{DineshDSouza}, \ttt{pastormarkburns}, \ttt{larryelder}, and \ttt{johncardillo}.  

An important limitation of our study is that Twitter users are not a representative sample of the general population \cite{mitchell2014}, and hashtag sampling introduces its own set of biases \cite{tufekci2014}. Moreover, differences in Twitter usage and propensity to tweet political content may also differ with political affiliation \cite{colleoni2014}. Consequently, it is important to compare our findings from Twitter to offline information. Our findings are consistent with a Quinnipiac poll that suggested that nearly one third of Republicans (but only 4\% of Democrats) considered counterprotesters to be more to blame than white supremacists for the violence at Charlottesville \cite{quinnipiac2017}. We observe that several of the communities on the Right that we obtained from modularity maximization of the \#Charlottesville retweet network also appear to reflect core participants of the `Unite the Right' rally \cite{fausset2017}, as indicated by the referencing by central nodes in these communities of white-supremacist hate symbols or influential personalities in the alt-right. 

Our investigation illustrated a stark distinction between Left and Right when we examined tweets that include the word `Trump', with criticism on the Left versus support on the Right (see Table \ref{table:wordcounts-sub}).  For example, the most numerous hashtags from the Left in tweets that include `Trump' were \#Impeachment and \#ImpeachTrump; by contrast, the most numerous hashtags were `\#Barcelona', `\#MAGA', and `\#fakenews' from accounts in communities on the Right. Our findings are consistent with the extreme polarization and political tribalism in United States society that have been described by other studies \cite{dimrock2014,iyengar2015,jacobson2017}. Such societal divisions are apparent on Twitter, as documented both by the present study and by prior ones \cite{colleoni2014,conover2011,feller2011,pennacchiotti2011}, including recent work that suggested that polarization on Twitter is increasing over time \cite{garimella2017}.   

It is also important to examine the role that fully automated accounts (`bots') and partially automated accounts (which have been dubbed `cyborgs' \cite{chu2012}) play in shaping conversations (especially political ones) on Twitter and other social-media platforms  \cite{beskow2018,bessi2016,chu2012,davis2016,freitas2015,stella2018,yardi2010}. Although an in-depth analysis of the role of bots in the \#Charlottesville discussion is beyond the scope of the present paper, it is likely that many bot accounts are present in our data set. For example, it has been noted that automated naming schemes are an indicator of bot accounts \cite{beskow2018}; and naming schemes that end in sequences of eight digits, as well as accounts that consist of hexadecimal strings, both occur in the \#Charlottesville data.  Detailed investigation of these accounts and their behavior is an important topic for future work. Sockpuppet accounts (i.e., false accounts that are operated by an entity \cite{yamak2018}), such as those that are operated by the Internet Research Agency in St. Petersburg, Russia \cite{house-intelligence2017,IRA_indictment}, can also play important roles in content propagation and thus warrant further investigation. Antipathy and distrust across party lines can provide opportunities for actors who seek to fan societal divisions. For example, our data set includes tweets by prominent accounts that were operated by the Internet Research Agency \cite{house-intelligence2017,IRA_indictment} and attacked both the Left and the Right.

Our approach of using media choice and PCA to characterize nodes is simple; it does not require labeled data, facilitating application to other fascinating topics. For example, it would be interesting to apply our approach to examine polarization on other topics (e.g., Brexit) and to see how polarization across political divides and attempts to bridge them change over time. It is not clear whether engagement with Twitter accounts with different viewpoints will decrease or increase polarization on divisive topics. For example, the empirical results of Bail et al. \cite{bail2018} suggest that exposure on Twitter to contrasting ideologies can lead to increased polarization. An interesting question is how exposure shapes viewpoints of individuals with `centrist' media preferences or ideologies. Our investigation focused primarily on the sign of a media PCA score, but the underlying media PCA score is continuous, and one can use it to examine media preferences in a more nuanced way. In particular, characterization of accounts with moderate (`centrist') media PCA scores, study of network structure and tweet content by these nodes, and tracking the evolution of these characteristics over time is both feasible and relevant. 

It is possible to refine our approach in various ways. For example, Landgraf and Lee \cite{landgraf2015} used a type of PCA that is built for Boolean data, and it may be insightful to compare our PCA results to ones that employ their approach. Examining additional PCA components besides the first one (on which we focused exclusively in the present paper) may also be helpful for understanding the diversity of interactions between media and other accounts on Twitter. Careful analysis of additional PCs, interpretations of them, and their relationship to network structure merit further investigation.  

Examining multiple ideological dimensions (e.g., as in studies of voting by legislators on bills \cite{voteview}) and simultaneous analysis of multiple types of Twitter relationships using the formalism of muiltilayer networks \cite{kivela2014} may also help deepen understanding of political communities and their news preferences after landmark news events. More broadly, we expect that our approach is generalizable to other contexts, and it may be helpful for examining other types of node characterizations (such as by analyzing different media outlets or types of followed accounts).  
Comparing network relationships and propagation of content across multiple social-media platforms is of particular interest, as amount, diversity, and characteristics of platform usage vary across different segments of the world's population. There have been studies of the propagation of memes \cite{zannettou2018b}, web addresses \cite{zannettou2019}, and anti-semitic content \cite{finkelstein2018} across different social-media platforms; and it is important to undertake further studies of linkages across networks and their effect on recruitment, content propagation, and public discourse.


\section*{List of abbreviations}

\begin{tabular}{ll}
API: & Application Program Interface \\
HITS: & Hyperlink-Induced Topic Search \\
PCA: & Principal Components Analysis \\
UTC: & Coordinated Universal Time \\
\end{tabular}


\section*{Declarations}

\paragraph{Acknowledgements.}
We thank Rob Bond, Heather Brooks, Matt Salganik, and Sam Zhang for helpful comments.


\paragraph{Availability of data and materials.}
We have made the de-identified data for the largest connected component of the retweet network, as well as code for analyzing these data, available (at \url{https://osf.io/487fw/}) via the Open Science Framework. For privacy reasons, we do not provide any account names; for similar reasons, we do not provide tweet content.

\paragraph{Competing interests.}
The authors declare that they have no competing interests.


\paragraph{Funding.}
The authors have no funding sources to acknowledge for this study.


\paragraph{Authors' contributions.}
JHT, MCE, STC, and MAP designed the study. JHT performed the analyses. JHT, MCE, STC, and MAP wrote the paper.  All authors read and approved the final manuscript.


\pagebreak




\newcommand{\BMCxmlcomment}[1]{}

\BMCxmlcomment{

<refgrp>

<bibl id="B1">
  <title><p>Far-right groups surge into national view in
  {Charlottesville}.</p></title>
  <aug>
    <au><snm>Fausset</snm><fnm>R.</fnm></au>
    <au><snm>Feuer</snm><fnm>A.</fnm></au>
  </aug>
  <source>New York Times</source>
  <pubdate>13 August 2017</pubdate>
  <note>Available at
  \url{https://www.nytimes.com/2017/08/13/us/far-right-groups-blaze-into-national-view-in-charlottesville.html}</note>
</bibl>

<bibl id="B2">
  <title><p>{Neo-Nazi} sympathizer pleads guilty to federal hate crimes for
  plowing car into protesters at {Charlottesville} rally.
  \url{https://www.washingtonpost.com/local/public-safety/neo-nazi-sympathizer-pleads-guilty-to-federal-hate-crimes-for-plowing-car-into-crowd-of-protesters-at-unite-the-right-rally-in-charlottesville/2019/03/27/2b947c32-50ab-11e9-8d28-f5149e5a2fda_story.html?utm_term=.738c2ac1e287}</p></title>
  <aug>
    <au><snm>Duggan</snm><fnm>P.</fnm></au>
    <au><snm>Jouvenal</snm><fnm>J.</fnm></au>
  </aug>
  <source>The Washington Post</source>
  <pubdate>27 March 2019</pubdate>
</bibl>

<bibl id="B3">
  <title><p>Trump defends initial remarks on {Charlottesville}: {A}gain blames
  `both sides'</p></title>
  <aug>
    <au><snm>Shear</snm><fnm>M. D.</fnm></au>
    <au><snm>Haberman</snm><fnm>M.</fnm></au>
  </aug>
  <source>New York Times</source>
  <pubdate>15 August 15 2017</pubdate>
  <note>Available at
  \url{https://www.nytimes.com/2017/08/15/us/politics/trump-press-conference-charlottesville.html}</note>
</bibl>

<bibl id="B4">
  <title><p>The algorithmic rise of the ``alt-right"</p></title>
  <aug>
    <au><snm>Daniels</snm><fnm>J.</fnm></au>
  </aug>
  <source>Contexts</source>
  <pubdate>2018</pubdate>
  <volume>17</volume>
  <issue>1</issue>
  <fpage>60</fpage>
  <lpage>65</lpage>
</bibl>

<bibl id="B5">
  <title><p>Big questions for social media {Big Data}: {R}epresentativeness,
  validity and other methodological pitfalls</p></title>
  <aug>
    <au><snm>Tufekci</snm><fnm>Z.</fnm></au>
  </aug>
  <source>Proceedings of the Eighth International AAAI Conference on Weblogs
  and Social Media</source>
  <pubdate>2014</pubdate>
  <fpage>505</fpage>
  <lpage>-514</lpage>
</bibl>

<bibl id="B6">
  <title><p>Social media usage: 2005-2015</p></title>
  <aug>
    <au><snm>Perrin</snm><fnm>A.</fnm></au>
  </aug>
  <pubdate>2015</pubdate>
  <note>Available at
  \url{http://www.pewinternet.org/2015/10/08/social-networking-usage-2005-2015/}</note>
</bibl>

<bibl id="B7">
  <title><p>Twitter and Tear Gas: {T}he Power and Fragility of Networked
  Protest</p></title>
  <aug>
    <au><snm>Tufekci</snm><fnm>Z.</fnm></au>
  </aug>
  <publisher>New Haven, CT, USA: Yale University Press</publisher>
  <pubdate>2017</pubdate>
</bibl>

<bibl id="B8">
  <title><p>Political polarization on {Twitter}</p></title>
  <aug>
    <au><snm>Conover</snm><fnm>M. D.</fnm></au>
    <au><snm>Ratkiewsicz</snm><fnm>J.</fnm></au>
    <au><snm>Francisco</snm><fnm>M.</fnm></au>
    <au><snm>Gon\c{c}alves</snm><fnm>B.</fnm></au>
    <au><snm>Flammini</snm><fnm>A.</fnm></au>
    <au><snm>Menczer</snm><fnm>F.</fnm></au>
  </aug>
  <source>Proceedings of the Fifth International AAAI Conference on Weblogs and
  Social Media</source>
  <pubdate>2011</pubdate>
  <fpage>89</fpage>
  <lpage>-96</lpage>
</bibl>

<bibl id="B9">
  <title><p>Influence and passivity in social media</p></title>
  <aug>
    <au><snm>Romero</snm><fnm>D. M.</fnm></au>
    <au><snm>Galuba</snm><fnm>W.</fnm></au>
    <au><snm>Asur</snm><fnm>S.</fnm></au>
    <au><snm>Huberman</snm><fnm>B. A.</fnm></au>
  </aug>
  <source>Proceedings of the 20th ACM International Conference on the World
  Wide Web -- Companion Volume (WWW'11 Companion)</source>
  <pubdate>2011</pubdate>
  <fpage>113</fpage>
  <lpage>114</lpage>
</bibl>

<bibl id="B10">
  <title><p>Echo chamber or public sphere? {Predicting} political orientation
  and measuring political homophily in {Twitter} using big data</p></title>
  <aug>
    <au><snm>Colleoni</snm><fnm>E.</fnm></au>
    <au><snm>Rozza</snm><fnm>A.</fnm></au>
    <au><snm>Arvidsson</snm><fnm>A.</fnm></au>
  </aug>
  <source>Journal of Communication</source>
  <pubdate>2014</pubdate>
  <volume>64</volume>
  <issue>2</issue>
  <fpage>317</fpage>
  <lpage>-332</lpage>
</bibl>

<bibl id="B11">
  <title><p>The Critical Periphery in the Growth of Social Protests</p></title>
  <aug>
    <au><snm>Barber{\'a}</snm><fnm>P</fnm></au>
    <au><snm>Wang</snm><fnm>N</fnm></au>
    <au><snm>Bonneau</snm><fnm>R</fnm></au>
    <au><snm>Jost</snm><fnm>JT</fnm></au>
    <au><snm>Nagler</snm><fnm>J</fnm></au>
    <au><snm>Tucker</snm><fnm>J</fnm></au>
    <au><snm>Gonz{\'a}lez Bail{\'o}n</snm><fnm>S</fnm></au>
  </aug>
  <source>PLoS ONE</source>
  <publisher>Public Library of Science</publisher>
  <pubdate>2015</pubdate>
  <volume>10</volume>
  <issue>11</issue>
  <fpage>e0143611</fpage>
  <url>https://doi.org/10.1371/journal.pone.0143611</url>
</bibl>

<bibl id="B12">
  <title><p>Interest communities and flow roles in directed networks: {T}he
  {T}witter network of the {UK} riots.</p></title>
  <aug>
    <au><snm>Beguerisse D{\'\i}az</snm><fnm>M.</fnm></au>
    <au><snm>Gardu{\~n}o Hern{\'a}ndez</snm><fnm>G</fnm></au>
    <au><snm>Vangelov</snm><fnm>B</fnm></au>
    <au><snm>Yaliraki</snm><fnm>SN</fnm></au>
    <au><snm>Barahona</snm><fnm>M</fnm></au>
  </aug>
  <source>Journal of the Royal Society: Interface</source>
  <pubdate>2014</pubdate>
  <volume>11</volume>
  <issue>101</issue>
</bibl>

<bibl id="B13">
  <title><p>A biased review of biases in {Twitter} studies on political
  collective action</p></title>
  <aug>
    <au><snm>Cihon</snm><fnm>P.</fnm></au>
    <au><snm>Yasseri</snm><fnm>T.</fnm></au>
  </aug>
  <source>Frontiers in Physics</source>
  <pubdate>2016</pubdate>
  <volume>4</volume>
  <fpage>34</fpage>
</bibl>

<bibl id="B14">
  <title><p>Beyond the hashtags: \#{Ferguson}, \#{Blacklivesmatter}, and the
  online struggle for offline justice</p></title>
  <aug>
    <au><snm>Freelon</snm><fnm>D.</fnm></au>
    <au><snm>{McIlwain}</snm><fnm>C. D.</fnm></au>
    <au><snm>Clar</snm><fnm>M.</fnm></au>
  </aug>
  <pubdate>2016</pubdate>
  <note>Available at \url{https://ssrn.com/abstract=2747066}</note>
</bibl>

<bibl id="B15">
  <title><p>Syria in the {Arab Spring}: {T}he integration of {Syria's} conflict
  with the {Arab} uprisings, 2011--2013</p></title>
  <aug>
    <au><snm>Lynch</snm><fnm>M.</fnm></au>
    <au><snm>Freelon</snm><fnm>D.</fnm></au>
    <au><snm>Aday</snm><fnm>S.</fnm></au>
  </aug>
  <source>Research and Politics</source>
  <pubdate>2014</pubdate>
  <volume>1</volume>
  <issue>3</issue>
  <fpage>2053168014549091</fpage>
</bibl>

<bibl id="B16">
  <title><p>Anatomy of Protest in the Digital Era: {A} Network Analysis of
  {Twitter} and {Occupy Wall Street}</p></title>
  <aug>
    <au><snm>Tremayne</snm><fnm>M.</fnm></au>
  </aug>
  <source>Social Movement Studies</source>
  <publisher>Routledge</publisher>
  <pubdate>2014</pubdate>
  <volume>13</volume>
  <issue>1</issue>
  <fpage>110</fpage>
  <lpage>126</lpage>
</bibl>

<bibl id="B17">
  <title><p>Social media and the decision to participate in political protest:
  {O}bservations from {Tahrir Square}</p></title>
  <aug>
    <au><snm>Tufekci</snm><fnm>Z.</fnm></au>
    <au><snm>Wilson</snm><fnm>C.</fnm></au>
  </aug>
  <source>Journal of Communication</source>
  <pubdate>2012</pubdate>
  <volume>62</volume>
  <issue>2</issue>
  <fpage>363</fpage>
  <lpage>379</lpage>
</bibl>

<bibl id="B18">
  <title><p>Pandemics in the Age of {Twitter}: Content Analysis of {Tweets}
  during the 2009 {H1N1} Outbreak</p></title>
  <aug>
    <au><snm>Chew</snm><fnm>C</fnm></au>
    <au><snm>Eysenbach</snm><fnm>G</fnm></au>
  </aug>
  <source>PLoS ONE</source>
  <publisher>Public Library of Science</publisher>
  <pubdate>2010</pubdate>
  <volume>5</volume>
  <issue>11</issue>
  <fpage>e14118</fpage>
  <url>http://dx.doi.org/10.1371
</bibl>

<bibl id="B19">
  <title><p>How to exploit {Twitter} for public health monitoring?</p></title>
  <aug>
    <au><snm>Denecke</snm><fnm>K.</fnm></au>
    <au><snm>Krieck</snm><fnm>M.</fnm></au>
    <au><snm>Otrusina</snm><fnm>L.</fnm></au>
    <au><snm>Smrz</snm><fnm>P.</fnm></au>
    <au><snm>Dolog</snm><fnm>P.</fnm></au>
    <au><snm>Nejdl</snm><fnm>W.</fnm></au>
    <au><snm>Velasco</snm><fnm>E.</fnm></au>
  </aug>
  <source>Methods of Information in Medicine</source>
  <pubdate>2013</pubdate>
  <volume>52</volume>
  <issue>4</issue>
  <fpage>326</fpage>
  <lpage>339</lpage>
</bibl>

<bibl id="B20">
  <title><p>Who Will Retweet This?: Automatically Identifying and Engaging
  Strangers on {Twitter} to Spread Information</p></title>
  <aug>
    <au><snm>Lee</snm><fnm>K</fnm></au>
    <au><snm>Mahmud</snm><fnm>J</fnm></au>
    <au><snm>Chen</snm><fnm>J</fnm></au>
    <au><snm>Zhou</snm><fnm>M</fnm></au>
    <au><snm>Nichols</snm><fnm>J</fnm></au>
  </aug>
  <source>Proceedings of the 19th International Conference on Intelligent User
  Interfaces</source>
  <publisher>New York, NY, USA: ACM</publisher>
  <series><title><p>IUI '14</p></title></series>
  <pubdate>2014</pubdate>
  <fpage>247</fpage>
  <lpage>-256</lpage>
</bibl>

<bibl id="B21">
  <title><p>Twitter influence on {UK} vaccination and antiviral uptake during
  the 2009 {H1N1} pandemic</p></title>
  <aug>
    <au><snm>{McNeill}</snm><fnm>A.</fnm></au>
    <au><snm>Harris</snm><fnm>P. R.</fnm></au>
    <au><snm>Briggs</snm><fnm>P.</fnm></au>
  </aug>
  <source>Frontiers in Public Health</source>
  <pubdate>2016</pubdate>
  <volume>4</volume>
  <fpage>26</fpage>
</bibl>

<bibl id="B22">
  <title><p>The dynamics of health behavior sentiments on a large online social
  network</p></title>
  <aug>
    <au><snm>Salath\'{e}</snm><fnm>M.</fnm></au>
    <au><snm>Vu</snm><fnm>D. W.</fnm></au>
    <au><snm>Khandelwal</snm><fnm>S.</fnm></au>
    <au><snm>Hunter</snm><fnm>D. R.</fnm></au>
  </aug>
  <source>EPJ Data Science</source>
  <pubdate>2013</pubdate>
  <volume>2</volume>
  <issue>1</issue>
  <fpage>4</fpage>
</bibl>

<bibl id="B23">
  <title><p>The use of {Twitter} to track levels of disease activity and public
  concern in the {US} during the influenza {A H1N1} pandemic</p></title>
  <aug>
    <au><snm>Signorini</snm><fnm>A.</fnm></au>
    <au><snm>Segre</snm><fnm>A. M.</fnm></au>
    <au><snm>Polgreen</snm><fnm>P. M.</fnm></au>
  </aug>
  <source>{PLoS ONE}</source>
  <pubdate>2011</pubdate>
  <volume>6</volume>
  <fpage>e19467</fpage>
</bibl>

<bibl id="B24">
  <title><p>Mass Media and the Contagion of Fear: The Case of {Ebola} in
  {America}</p></title>
  <aug>
    <au><snm>Towers</snm><fnm>S</fnm></au>
    <au><snm>Afzal</snm><fnm>S</fnm></au>
    <au><snm>Bernal</snm><fnm>G</fnm></au>
    <au><snm>Bliss</snm><fnm>N</fnm></au>
    <au><snm>Brown</snm><fnm>S</fnm></au>
    <au><snm>Espinoza</snm><fnm>B</fnm></au>
    <au><snm>Jackson</snm><fnm>J</fnm></au>
    <au><snm>Judson Garcia</snm><fnm>J</fnm></au>
    <au><snm>Khan</snm><fnm>M</fnm></au>
    <au><snm>Lin</snm><fnm>M</fnm></au>
    <au><snm>Mamada</snm><fnm>R</fnm></au>
    <au><snm>Moreno</snm><fnm>VM</fnm></au>
    <au><snm>Nazari</snm><fnm>F</fnm></au>
    <au><snm>Okuneye</snm><fnm>K</fnm></au>
    <au><snm>Ross</snm><fnm>ML</fnm></au>
    <au><snm>Rodriguez</snm><fnm>C</fnm></au>
    <au><snm>Medlock</snm><fnm>J</fnm></au>
    <au><snm>Ebert</snm><fnm>D</fnm></au>
    <au><snm>Castillo Chavez</snm><fnm>C</fnm></au>
  </aug>
  <source>PLoS ONE</source>
  <publisher>Public Library of Science</publisher>
  <pubdate>2015</pubdate>
  <volume>10</volume>
  <issue>6</issue>
  <fpage>1</fpage>
  <lpage>13</lpage>
  <url>http://dx.doi.org/10.1371
</bibl>

<bibl id="B25">
  <title><p>Information contagion: {A}n empirical study of the spread of news
  on {Digg} and {Twitter} social networks</p></title>
  <aug>
    <au><snm>{Lerman}</snm><fnm>K.</fnm></au>
    <au><snm>{Ghosh}</snm><fnm>R.</fnm></au>
  </aug>
  <source>ICWSM</source>
  <pubdate>2010</pubdate>
  <volume>10</volume>
  <fpage>90</fpage>
  <lpage>97</lpage>
</bibl>

<bibl id="B26">
  <title><p>Virality Prediction and Community Structure in Social
  Networks</p></title>
  <aug>
    <au><snm>Weng</snm><fnm>L</fnm></au>
    <au><snm>Menczer</snm><fnm>F</fnm></au>
    <au><snm>Ahn</snm><fnm>YY</fnm></au>
  </aug>
  <source>Scientific Reports</source>
  <publisher>The Author(s) SN -</publisher>
  <pubdate>2013</pubdate>
  <volume>3</volume>
  <fpage>2522</fpage>
  <url>http://dx.doi.org/10.1038/srep02522</url>
</bibl>

<bibl id="B27">
  <title><p>Online extremism and the communities that sustain it: {D}etecting
  the {ISIS} supporting community on {Twitter}</p></title>
  <aug>
    <au><snm>Benigni</snm><fnm>M. C.</fnm></au>
    <au><snm>Joseph</snm><fnm>K.</fnm></au>
    <au><snm>Carley</snm><fnm>K. M.</fnm></au>
  </aug>
  <source>{PLoS ONE}</source>
  <pubdate>2017</pubdate>
  <volume>12</volume>
  <issue>12</issue>
  <fpage>e0181405</fpage>
</bibl>

<bibl id="B28">
  <title><p>A Long-Term Analysis of Polarization on {Twitter}</p></title>
  <aug>
    <au><snm>Garimella</snm><fnm>K</fnm></au>
    <au><snm>Weber</snm><fnm>I</fnm></au>
  </aug>
  <source>International AAAI Conference on Web and Social Media</source>
  <pubdate>2017</pubdate>
  <fpage>528</fpage>
  <lpage>531</lpage>
</bibl>

<bibl id="B29">
  <title><p>Measuring political polarization: {Twitter} shows the two sides of
  {Venezuela}</p></title>
  <aug>
    <au><snm>Morales</snm><fnm>A. J.</fnm></au>
    <au><snm>Borondo</snm><fnm>J.</fnm></au>
    <au><snm>Losada</snm><fnm>J. C.</fnm></au>
    <au><snm>Benito</snm><fnm>R. M.</fnm></au>
  </aug>
  <source>Chaos</source>
  <pubdate>2015</pubdate>
  <volume>25</volume>
  <issue>3</issue>
  <fpage>033114</fpage>
</bibl>

<bibl id="B30">
  <title><p>Twitter reaction to events often at odds with overall public
  opinion</p></title>
  <aug>
    <au><snm>Mitchell</snm><fnm>A.</fnm></au>
    <au><snm>Hitlin</snm><fnm>P.</fnm></au>
  </aug>
  <pubdate>2014</pubdate>
  <note>Available at
  \url{http://www.pewresearch.org/2013/03/04/twitter-reaction-to-
  events-often-at-odds-with-overall-public-opinion/}</note>
</bibl>

<bibl id="B31">
  <title><p>Is the sample good enough? {Comparing} data from {Twitter's}
  streaming {API} with {Twitter's} firehose</p></title>
  <aug>
    <au><snm>Morstatter</snm><fnm>F.</fnm></au>
    <au><snm>Pfeffer</snm><fnm>J.</fnm></au>
    <au><snm>Liu</snm><fnm>H.</fnm></au>
    <au><snm>Carley</snm><fnm>K. M.</fnm></au>
  </aug>
  <source>Proceedings of the Seventh International AAAI Conference on Weblogs
  and Social Media</source>
  <pubdate>2014</pubdate>
  <fpage>400</fpage>
  <lpage>-408</lpage>
</bibl>

<bibl id="B32">
  <title><p>Exposure to opposing views on social media can increase political
  polarization</p></title>
  <aug>
    <au><snm>Bail</snm><fnm>C. A.</fnm></au>
    <au><snm>Argyle</snm><fnm>L. P.</fnm></au>
    <au><snm>Brown</snm><fnm>T. W.</fnm></au>
    <au><snm>Bumpus</snm><fnm>J. P.</fnm></au>
    <au><snm>Chen</snm><fnm>H.</fnm></au>
    <au><snm>{Fallin Hunzaker}</snm><fnm>M. B.</fnm></au>
    <au><snm>Lee</snm><fnm>J.</fnm></au>
    <au><snm>Mann</snm><fnm>M.</fnm></au>
    <au><snm>Merhout</snm><fnm>F.</fnm></au>
    <au><snm>Volfovsky</snm><fnm>A.</fnm></au>
  </aug>
  <source>Proceedings of the National Academy of Sciences of the United States
  of America</source>
  <pubdate>2018</pubdate>
  <volume>115</volume>
  <issue>37</issue>
  <fpage>9216</fpage>
  <lpage>9221</lpage>
</bibl>

<bibl id="B33">
  <title><p>Filter bubbles, echo chambers, and online news
  consumption</p></title>
  <aug>
    <au><snm>Flaxman</snm><fnm>S.</fnm></au>
    <au><snm>Goel</snm><fnm>S.</fnm></au>
    <au><snm>Rao</snm><fnm>J. M.</fnm></au>
  </aug>
  <source>Public Opinion Quarterly</source>
  <pubdate>2016</pubdate>
  <volume>80</volume>
  <fpage>298</fpage>
  <lpage>-320</lpage>
</bibl>

<bibl id="B34">
  <title><p>{Republic.com}</p></title>
  <aug>
    <au><snm>Sunstein</snm><fnm>C. R.</fnm></au>
  </aug>
  <publisher>Princeton, NJ, USA: Princeton University Press</publisher>
  <pubdate>2001</pubdate>
</bibl>

<bibl id="B35">
  <title><p>Online groups and political discourse: {D}o online discussion
  spaces facilitate exposure to political disagreement?</p></title>
  <aug>
    <au><snm>Wojcieszak</snm><fnm>M.</fnm></au>
    <au><snm>Mutz</snm><fnm>D.</fnm></au>
  </aug>
  <source>Journal of Communication</source>
  <pubdate>2009</pubdate>
  <volume>59</volume>
  <fpage>40</fpage>
  <lpage>-56</lpage>
</bibl>

<bibl id="B36">
  <title><p>Social media and fake news in the 2016 election</p></title>
  <aug>
    <au><snm>Allcott</snm><fnm>H.</fnm></au>
    <au><snm>Gentzkow</snm><fnm>M.</fnm></au>
  </aug>
  <source>Journal of Economic Perspectives</source>
  <pubdate>2017</pubdate>
  <volume>31</volume>
  <issue>2</issue>
  <fpage>211</fpage>
  <lpage>236</lpage>
</bibl>

<bibl id="B37">
  <title><p>The Filter Bubble: {W}hat the {Internet} is Hiding from
  You</p></title>
  <aug>
    <au><snm>Pariser</snm><fnm>E.</fnm></au>
  </aug>
  <publisher>London, UK: Penguin Random House UK</publisher>
  <pubdate>2011</pubdate>
</bibl>

<bibl id="B38">
  <title><p>Divided they tweet: {T}he network structure of political
  microbloggers and discussion topics</p></title>
  <aug>
    <au><snm>Feller</snm><fnm>A.</fnm></au>
    <au><snm>Kuhnert</snm><fnm>M.</fnm></au>
    <au><snm>Sprenger</snm><fnm>T. O.</fnm></au>
    <au><snm>Welpe</snm><fnm>I.</fnm></au>
  </aug>
  <source>Proceedings of the fifth international AAAI conference on weblogs and
  social media (ICWSM 11)</source>
  <pubdate>2011</pubdate>
  <fpage>474</fpage>
  <lpage>477</lpage>
</bibl>

<bibl id="B39">
  <title><p>A machine learning approach to {Twitter} user
  classification</p></title>
  <aug>
    <au><snm>Pennacchiotti</snm><fnm>M.</fnm></au>
    <au><snm>Popescu</snm><fnm>A.</fnm></au>
  </aug>
  <source>Proceedings of AAAI conference on weblogs and social media (ICWSM
  2011)</source>
  <pubdate>2011</pubdate>
  <fpage>281</fpage>
  <lpage>288</lpage>
</bibl>

<bibl id="B40">
  <title><p>Principal Components Analysis</p></title>
  <aug>
    <au><snm>Jolliffe</snm><fnm>I. T.</fnm></au>
  </aug>
  <publisher>Heidelberg, Germany: Springer-Verlag</publisher>
  <edition>2</edition>
  <series><title><p>Springer Series in Statistics</p></title></series>
  <pubdate>2002</pubdate>
</bibl>

<bibl id="B41">
  <title><p>Fair and balanced? {Q}uantifying media bias through crowdsourced
  content analysis</p></title>
  <aug>
    <au><snm>Budak</snm><fnm>C.</fnm></au>
    <au><snm>Goel</snm><fnm>S.</fnm></au>
    <au><snm>Rao</snm><fnm>J. M.</fnm></au>
  </aug>
  <source>Public Opinion Quarterly</source>
  <pubdate>2016</pubdate>
  <volume>80</volume>
  <issue>S1</issue>
  <fpage>250</fpage>
  <lpage>-271</lpage>
</bibl>

<bibl id="B42">
  <title><p>Political polarization &amp media habits</p></title>
  <aug>
    <au><snm>Mitchell</snm><fnm>A.</fnm></au>
    <au><snm>Gottfried</snm><fnm>J.</fnm></au>
    <au><snm>Kiley</snm><fnm>J.</fnm></au>
    <au><snm>Masa</snm><fnm>K. E.</fnm></au>
  </aug>
  <pubdate>2014</pubdate>
  <note>Available at
  \url{http://www.journalism.org/2014/10/21/political-polarization-media-habits/}</note>
</bibl>

<bibl id="B43">
  <title><p>Measuring user influence in {Twitter}: {T}he million follower
  fallacy</p></title>
  <aug>
    <au><snm>Cha</snm><fnm>M.</fnm></au>
    <au><snm>Haddadi</snm><fnm>H.</fnm></au>
    <au><snm>Benevenuto</snm><fnm>F.</fnm></au>
    <au><snm>Gummadi</snm><fnm>K. P.</fnm></au>
  </aug>
  <source>4th International AAAI Conference on Weblogs and Social Media
  (ICWSM)</source>
  <pubdate>2010</pubdate>
  <fpage>10</fpage>
  <lpage>17</lpage>
</bibl>

<bibl id="B44">
  <title><p>Networks</p></title>
  <aug>
    <au><snm>Newman</snm><fnm>M. E. J.</fnm></au>
  </aug>
  <publisher>Oxford: Oxford University Press</publisher>
  <edition>Second</edition>
  <pubdate>2018</pubdate>
</bibl>

<bibl id="B45">
  <title><p>The anatomy of a large-scale hypertextual {Web} search
  engine</p></title>
  <aug>
    <au><snm>Brin</snm><fnm>S.</fnm></au>
    <au><snm>Page</snm><fnm>L.</fnm></au>
  </aug>
  <source>Computer Networks and ISDN Systems</source>
  <pubdate>1998</pubdate>
  <volume>30</volume>
  <fpage>107</fpage>
  <lpage>117</lpage>
</bibl>

<bibl id="B46">
  <title><p>A set of measures of centrality based on betweenness</p></title>
  <aug>
    <au><snm>Freeman</snm><fnm>L.</fnm></au>
  </aug>
  <source>Sociometry</source>
  <pubdate>1977</pubdate>
  <volume>40</volume>
  <issue>1</issue>
  <fpage>35</fpage>
  <lpage>-41</lpage>
</bibl>

<bibl id="B47">
  <title><p>Authoritative sources in a hyperlinked environment</p></title>
  <aug>
    <au><snm>Kleinberg</snm><fnm>J. M.</fnm></au>
  </aug>
  <source>Journal of the Association for Computing Machinery</source>
  <pubdate>1999</pubdate>
  <volume>46</volume>
  <fpage>604</fpage>
  <lpage>632</lpage>
</bibl>

<bibl id="B48">
  <title><p>{BotOrNot}: {A} system to evaluate social bots</p></title>
  <aug>
    <au><snm>Davis</snm><fnm>C. A.</fnm></au>
    <au><snm>Varol</snm><fnm>O.</fnm></au>
    <au><snm>Ferrara</snm><fnm>E.</fnm></au>
    <au><snm>Flammini</snm><fnm>A.</fnm></au>
    <au><snm>Menczer</snm><fnm>F.</fnm></au>
  </aug>
  <source>Proceedings of the 25th International World Wide Web Conference
  Companion</source>
  <pubdate>2016</pubdate>
  <fpage>273</fpage>
  <lpage>274</lpage>
</bibl>

<bibl id="B49">
  <title><p>Identifying authorities in online communities</p></title>
  <aug>
    <au><snm>Bouguessa</snm><fnm>M.</fnm></au>
    <au><snm>{Ben Romdhane}</snm><fnm>L.</fnm></au>
  </aug>
  <source>ACM Transactions on Intelligent Systems and Technology</source>
  <pubdate>2015</pubdate>
  <volume>6</volume>
  <issue>3</issue>
  <fpage>30</fpage>
</bibl>

<bibl id="B50">
  <title><p>{TwitterRank}: finding topic-sensitive influential
  twitterers</p></title>
  <aug>
    <au><snm>Weng</snm><fnm>J.</fnm></au>
    <au><snm>Lim</snm><fnm>E. P.</fnm></au>
    <au><snm>Jiang</snm><fnm>J.</fnm></au>
    <au><snm>He</snm><fnm>Q.</fnm></au>
  </aug>
  <source>Proceedings of the 3rd ACM International Conference on Web Search and
  Data Mining (WSDM'10)</source>
  <pubdate>2010</pubdate>
  <fpage>261</fpage>
  <lpage>270</lpage>
</bibl>

<bibl id="B51">
  <title><p>Communities in networks</p></title>
  <aug>
    <au><snm>Porter</snm><fnm>M. A.</fnm></au>
    <au><snm>Onnela</snm><fnm>J. P.</fnm></au>
    <au><snm>Mucha</snm><fnm>P. J.</fnm></au>
  </aug>
  <source>Notices of the American Mathematical Society</source>
  <pubdate>2009</pubdate>
  <volume>56</volume>
  <issue>9</issue>
  <fpage>1082</fpage>
  <lpage>-1097</lpage>
  <note>Second page range: 1164--1166</note>
</bibl>

<bibl id="B52">
  <title><p>Community detection in networks: {A} user guide</p></title>
  <aug>
    <au><snm>Fortunato</snm><fnm>S</fnm></au>
    <au><snm>Hric</snm><fnm>D</fnm></au>
  </aug>
  <source>Phys. Rep.</source>
  <pubdate>2016</pubdate>
  <volume>659</volume>
  <fpage>1</fpage>
  <lpage>-44</lpage>
</bibl>

<bibl id="B53">
  <title><p>Think locally, act locally: {D}etection of small, medium-sized, and
  large communities in large networks</p></title>
  <aug>
    <au><snm>Jeub</snm><fnm>L. G. S.</fnm></au>
    <au><snm>Balachandran</snm><fnm>P.</fnm></au>
    <au><snm>Porter</snm><fnm>M. A.</fnm></au>
    <au><snm>Mucha</snm><fnm>P. J.</fnm></au>
    <au><snm>Mahoney</snm><fnm>M. W.</fnm></au>
  </aug>
  <source>Physical Review E</source>
  <pubdate>2015</pubdate>
  <volume>91</volume>
  <issue>1</issue>
  <fpage>012821</fpage>
</bibl>

<bibl id="B54">
  <title><p>Dynamics and control of diseases in networks with community
  structure</p></title>
  <aug>
    <au><snm>Salath\'{e}</snm><fnm>M.</fnm></au>
    <au><snm>Jones</snm><fnm>J. H.</fnm></au>
  </aug>
  <source>{PLoS Computational Biology}</source>
  <pubdate>2010</pubdate>
  <volume>6</volume>
  <issue>4</issue>
  <fpage>e1000736</fpage>
</bibl>

<bibl id="B55">
  <title><p>Word usage mirrors community structure in the online social network
  {Twitter}</p></title>
  <aug>
    <au><snm>Bryden</snm><fnm>J.</fnm></au>
    <au><snm>Funk</snm><fnm>S.</fnm></au>
    <au><snm>Jansen</snm><fnm>V. A. A.</fnm></au>
  </aug>
  <source>EPJ Data Science</source>
  <pubdate>2013</pubdate>
  <volume>2</volume>
  <issue>1</issue>
  <fpage>3</fpage>
</bibl>

<bibl id="B56">
  <title><p>Quantifying political leaning from tweets, retweets, and
  retweeters</p></title>
  <aug>
    <au><snm>Wong</snm><fnm>F. M. F.</fnm></au>
    <au><snm>Tan</snm><fnm>C. W.</fnm></au>
    <au><snm>Sen</snm><fnm>S.</fnm></au>
    <au><snm>Chiang</snm><fnm>M.</fnm></au>
  </aug>
  <source>IEEE Transactions on Knowledge and Data Engineering</source>
  <pubdate>2016</pubdate>
  <volume>28</volume>
  <issue>8</issue>
  <fpage>2158</fpage>
  <lpage>2172</lpage>
</bibl>

<bibl id="B57">
  <title><p>Challenging racist nativist framing: Acknowledging the community
  cultural wealth of undocumented {Chicana} college students to reframe the
  immigration debate</p></title>
  <aug>
    <au><snm>Huber</snm><fnm>L. P.</fnm></au>
  </aug>
  <source>Harvard Educational Review</source>
  <pubdate>2009</pubdate>
  <volume>79</volume>
  <issue>4</issue>
  <fpage>704</fpage>
  <lpage>730</lpage>
</bibl>

<bibl id="B58">
  <title><p>{The Framing of Immigration}. {Available} at
  \url{https://escholarship.org/uc/item/0j89f85g}</p></title>
  <aug>
    <au><snm>Lakoff</snm><fnm>G.</fnm></au>
    <au><snm>Ferguson</snm><fnm>S.</fnm></au>
  </aug>
  <source>UC Berkeley</source>
  <pubdate>2006</pubdate>
</bibl>

<bibl id="B59">
  <title><p>How the `alt-right' came to dominate the comments on {Trump's
  Facebook} page</p></title>
  <aug>
    <au><snm>Morgan</snm><fnm>J.</fnm></au>
  </aug>
  <source>The Atlantic</source>
  <pubdate>2017</pubdate>
  <note>Available at
  \url{https://www.theatlantic.com/politics/archive/2017/01/how-the-alt-right-influenced-trump-supporters-language-on-facebook/513593/}</note>
</bibl>

<bibl id="B60">
  <title><p>Bit by Bit: {S}ocial Research in the Digital Age</p></title>
  <aug>
    <au><snm>Salganik</snm><fnm>M. J.</fnm></au>
  </aug>
  <publisher>Princeton, NJ, USA: Princeton University Press</publisher>
  <pubdate>2017</pubdate>
</bibl>

<bibl id="B61">
  <title><p>About verified accounts</p></title>
  <aug>
    <au><cnm>{Twitter, Inc.}</cnm></au>
  </aug>
  <source>\url{https://help.twitter.com/en/managing-your-account/about-twitter-verified-accounts}</source>
  <pubdate>2019. Last accessed: May 8, 2019.</pubdate>
</bibl>

<bibl id="B62">
  <title><p>Dimensionality reduction for binary data through the projection of
  natural parameters</p></title>
  <aug>
    <au><snm>Landgraf</snm><fnm>A. J.</fnm></au>
    <au><snm>Lee</snm><fnm>Y.</fnm></au>
  </aug>
  <source>arXiv</source>
  <pubdate>2015</pubdate>
  <volume>arXiv:1510.06112</volume>
</bibl>

<bibl id="B63">
  <title><p>Ideological segregation online and offline</p></title>
  <aug>
    <au><snm>Gentzkow</snm><fnm>M.</fnm></au>
    <au><snm>Shapiro</snm><fnm>J. M.</fnm></au>
  </aug>
  <source>Quarterly Journal of Economics</source>
  <pubdate>2011</pubdate>
  <volume>126</volume>
  <fpage>1799</fpage>
  <lpage>1839</lpage>
</bibl>

<bibl id="B64">
  <title><p>A measure of media bias</p></title>
  <aug>
    <au><snm>Groseclose</snm><fnm>T.</fnm></au>
    <au><snm>Milyo</snm><fnm>J.</fnm></au>
  </aug>
  <source>The Quarterly Journal of Economics</source>
  <pubdate>2005</pubdate>
  <volume>120</volume>
  <issue>4</issue>
  <fpage>1191</fpage>
  <lpage>1237</lpage>
</bibl>

<bibl id="B65">
  <title><p>Media Bias Chart: Version 4.0</p></title>
  <aug>
    <au><snm>Otero</snm><fnm>V</fnm></au>
  </aug>
  <pubdate>2018</pubdate>
  <note>Available at \url{https://www.adfontesmedia.com}</note>
</bibl>

<bibl id="B66">
  <title><p>Social network analysis for social neuroscientists</p></title>
  <aug>
    <au><snm>Baek</snm><fnm>E. C.</fnm></au>
    <au><snm>Porter</snm><fnm>M. A.</fnm></au>
    <au><snm>Parkinson</snm><fnm>C.</fnm></au>
  </aug>
  <source>arXiv</source>
  <pubdate>2019</pubdate>
  <volume>arXiv:1909.11894</volume>
</bibl>

<bibl id="B67">
  <title><p>{PageRank} beyond the {W}eb</p></title>
  <aug>
    <au><snm>Gleich</snm><fnm>D. F.</fnm></au>
  </aug>
  <source>SIAM Review</source>
  <pubdate>2015</pubdate>
  <volume>57</volume>
  <issue>3</issue>
  <fpage>321</fpage>
  <lpage>-363</lpage>
</bibl>

<bibl id="B68">
  <title><p>Fast unfolding of communities in large networks</p></title>
  <aug>
    <au><snm>Blondel</snm><fnm>VD</fnm></au>
    <au><snm>Guillaume</snm><fnm>JL</fnm></au>
    <au><snm>Lambiotte</snm><fnm>R</fnm></au>
    <au><snm>Lefebvre</snm><fnm>E</fnm></au>
  </aug>
  <source>Journal of Statistical Mechanics: Theory and Experiment</source>
  <pubdate>2008</pubdate>
  <volume>2008</volume>
  <issue>10</issue>
  <fpage>P10008</fpage>
  <url>http://stacks.iop.org/1742-5468/2008/i=10/a=P10008</url>
</bibl>

<bibl id="B69">
  <title><p>Finding and evaluating community structure in networks</p></title>
  <aug>
    <au><snm>Newman</snm><fnm>M. E. J.</fnm></au>
    <au><snm>Girvan</snm><fnm>M.</fnm></au>
  </aug>
  <source>Physical Review E</source>
  <pubdate>2004</pubdate>
  <volume>69</volume>
  <fpage>026113</fpage>
</bibl>

<bibl id="B70">
  <title><p>Finding community structure in networks using the eigenvectors of
  matrices</p></title>
  <aug>
    <au><snm>Newman</snm><fnm>M. E. J.</fnm></au>
  </aug>
  <source>Physical Review E</source>
  <pubdate>2006</pubdate>
  <volume>74</volume>
  <fpage>036104</fpage>
</bibl>

<bibl id="B71">
  <title><p>Maps of random walks on complex networks reveal community
  structure</p></title>
  <aug>
    <au><snm>Rosvall</snm><fnm>M.</fnm></au>
    <au><snm>Bergstrom</snm><fnm>C. T.</fnm></au>
  </aug>
  <source>Proceedings of the National Academy of Sciences of the United States
  of America</source>
  <pubdate>2008</pubdate>
  <volume>105</volume>
  <issue>4</issue>
  <fpage>1118</fpage>
  <lpage>1123</lpage>
</bibl>

<bibl id="B72">
  <title><p>Bayesian stochastic blockmodeling</p></title>
  <aug>
    <au><snm>Peixoto</snm><fnm>T. P.</fnm></au>
  </aug>
  <source>arXiv:1705.10225</source>
  <pubdate>2017</pubdate>
  <note>Chapter in ``Advances in Network Clustering and Blockmodeling'', edited
  by P. Doreian, V. Batagelj, A. Ferligoj, (John Wiley \& Sons, New York City,
  USA [forthcoming]).</note>
</bibl>

<bibl id="B73">
  <title><p>Size reduction of complex networks preserving
  modularity</p></title>
  <aug>
    <au><snm>Arenas</snm><fnm>A</fnm></au>
    <au><snm>Duch</snm><fnm>J</fnm></au>
    <au><snm>Fern{\'a}ndez</snm><fnm>A</fnm></au>
    <au><snm>G{\'o}mez</snm><fnm>S</fnm></au>
  </aug>
  <source>New Journal of Physics</source>
  <pubdate>2007</pubdate>
  <volume>9</volume>
  <issue>6</issue>
  <fpage>176</fpage>
  <url>http://stacks.iop.org/1367-2630/9/i=6/a=176</url>
</bibl>

<bibl id="B74">
  <title><p>Community structure in directed networks</p></title>
  <aug>
    <au><snm>Leicht</snm><fnm>E. A.</fnm></au>
    <au><snm>Newman</snm><fnm>M. E. J.</fnm></au>
  </aug>
  <source>Physics Review Letters</source>
  <pubdate>2008</pubdate>
  <volume>100</volume>
  <fpage>118703</fpage>
</bibl>

<bibl id="B75">
  <title><p>Random walks, {M}arkov processes and the multiscale modular
  organization of complex networks</p></title>
  <aug>
    <au><snm>Lambiotte</snm><fnm>R.</fnm></au>
    <au><snm>Delvenne</snm><fnm>J. C.</fnm></au>
    <au><snm>Barahona</snm><fnm>M.</fnm></au>
  </aug>
  <source>IEEE Transactions on Network Science and Engineering</source>
  <pubdate>2015</pubdate>
  <volume>1</volume>
  <issue>2</issue>
  <fpage>76</fpage>
  <lpage>-90</lpage>
</bibl>

<bibl id="B76">
  <title><p>Configuring random graph models with fixed degree
  sequences</p></title>
  <aug>
    <au><snm>Fosdick</snm><fnm>B. K.</fnm></au>
    <au><snm>Larremore</snm><fnm>D. B.</fnm></au>
    <au><snm>Nishimura</snm><fnm>J.</fnm></au>
    <au><snm>Ugander</snm><fnm>J.</fnm></au>
  </aug>
  <source>SIAM Review</source>
  <pubdate>2018</pubdate>
  <volume>60</volume>
  <issue>2</issue>
  <fpage>315</fpage>
  <lpage>-355</lpage>
</bibl>

<bibl id="B77">
  <title><p>A generalized {L}ouvain method for community detection implemented
  in {{\sc Matlab}}</p></title>
  <aug>
    <au><snm>Jeub</snm><fnm>LGS</fnm></au>
    <au><snm>Bazzi</snm><fnm>M</fnm></au>
    <au><snm>Jutla</snm><fnm>IS</fnm></au>
    <au><snm>Mucha</snm><fnm>PJ</fnm></au>
  </aug>
  <source>Available at \url{http://netwiki.amath.unc.edu/GenLouvain}</source>
  <pubdate>2011--2016</pubdate>
  <note>Version 2.0</note>
</bibl>

<bibl id="B78">
  <title><p>Community structure in time-dependent, multiscale, and multiplex
  networks</p></title>
  <aug>
    <au><snm>Mucha</snm><fnm>P. J.</fnm></au>
    <au><snm>Richardson</snm><fnm>T.</fnm></au>
    <au><snm>Macon</snm><fnm>K.</fnm></au>
    <au><snm>Porter</snm><fnm>M. A.</fnm></au>
    <au><snm>Onnela</snm><fnm>J. P.</fnm></au>
  </aug>
  <source>Science</source>
  <pubdate>2010</pubdate>
  <volume>328</volume>
  <issue>5980</issue>
  <fpage>876</fpage>
  <lpage>878</lpage>
</bibl>

<bibl id="B79">
  <title><p>The {MapEquation} software package, available online at
  \url{http://www.mapequation.org}</p></title>
  <aug>
    <au><snm>Edler</snm><fnm>D.</fnm></au>
    <au><snm>Rosvall</snm><fnm>M.</fnm></au>
  </aug>
  <pubdate>Last accessed: February 13, 2019</pubdate>
</bibl>

<bibl id="B80">
  <title><p>A mathematical theory of communication</p></title>
  <aug>
    <au><snm>Shannon</snm><fnm>C. E.</fnm></au>
  </aug>
  <source>The Bell System Technical Journal</source>
  <pubdate>1948</pubdate>
  <volume>27</volume>
  <fpage>379</fpage>
  <lpage>423and623-656</lpage>
</bibl>

<bibl id="B81">
  <title><p>Trump's most influential white nationalist troll is a {Middlebury}
  grad who lives in {Manhattan}. {Available at: }
  \url{https://www.huffpost.com/entry/trump-white-nationalist-troll-ricky-vaughn_n_5ac53167e4b09ef3b2432627}</p></title>
  <aug>
    <au><snm>{O'Brien}</snm><fnm>L.</fnm></au>
  </aug>
  <source>The Huffington Post</source>
  <pubdate>2018</pubdate>
</bibl>

<bibl id="B82">
  <title><p>Performance of modularity maximization in practical
  contexts</p></title>
  <aug>
    <au><snm>Good</snm><fnm>B. H.</fnm></au>
    <au><snm>{de Montjoye}</snm><fnm>{Y.-V.}</fnm></au>
    <au><snm>Clauset</snm><fnm>A.</fnm></au>
  </aug>
  <source>Physical Review E</source>
  <pubdate>2010</pubdate>
  <volume>81</volume>
  <fpage>046106</fpage>
</bibl>

<bibl id="B83">
  <title><p>Resolution limit in community detection</p></title>
  <aug>
    <au><snm>Fortunato</snm><fnm>S</fnm></au>
    <au><snm>Barth{\'e}lemy</snm><fnm>M</fnm></au>
  </aug>
  <source>Proceedings of the National Academy of Sciences of the United States
  of America</source>
  <publisher>National Academy of Sciences</publisher>
  <pubdate>2007</pubdate>
  <volume>104</volume>
  <issue>1</issue>
  <fpage>36</fpage>
  <lpage>-41</lpage>
</bibl>

<bibl id="B84">
  <title><p>Centrality in modular networks</p></title>
  <aug>
    <au><snm>Ghalmane</snm><fnm>Z.</fnm></au>
    <au><snm>Hassouni</snm><fnm>M. E.</fnm></au>
    <au><snm>Cherifi</snm><fnm>C.</fnm></au>
    <au><snm>Cherifi</snm><fnm>H.</fnm></au>
  </aug>
  <source>EPJ Data Science</source>
  <pubdate>2019</pubdate>
  <volume>8</volume>
  <issue>15</issue>
</bibl>

<bibl id="B85">
  <title><p>Functional cartography of complex metabolic networks</p></title>
  <aug>
    <au><snm>Guimer\`{a}</snm><fnm>R.</fnm></au>
    <au><snm>Amaral</snm><fnm>L. A. N.</fnm></au>
  </aug>
  <source>Nature</source>
  <pubdate>2005</pubdate>
  <volume>433</volume>
  <fpage>895</fpage>
  <lpage>-900</lpage>
</bibl>

<bibl id="B86">
  <title><p>Dynamical clustering of exchange rates</p></title>
  <aug>
    <au><snm>Fenn</snm><fnm>D. J.</fnm></au>
    <au><snm>Porter</snm><fnm>M. A.</fnm></au>
    <au><snm>Mucha</snm><fnm>P. J.</fnm></au>
    <au><snm>McDonald</snm><fnm>M.</fnm></au>
    <au><snm>Williams</snm><fnm>S.</fnm></au>
    <au><snm>Johnson</snm><fnm>N. F.</fnm></au>
    <au><snm>Jones</snm><fnm>N. S.</fnm></au>
  </aug>
  <source>Quantitative Finance</source>
  <publisher>Taylor \& Francis</publisher>
  <pubdate>2012</pubdate>
  <volume>12</volume>
  <issue>10</issue>
  <fpage>1493</fpage>
  <lpage>-1520</lpage>
</bibl>

<bibl id="B87">
  <title><p>Distribution of node characteristics in complex
  networks</p></title>
  <aug>
    <au><snm>Park</snm><fnm>J.</fnm></au>
    <au><snm>Barab\'asi</snm><fnm>A. L.</fnm></au>
  </aug>
  <source>Proceedings of the National Academy of Sciences of the United States
  of America</source>
  <pubdate>2007</pubdate>
  <volume>104</volume>
  <issue>46</issue>
  <fpage>17916</fpage>
  <lpage>-17920</lpage>
</bibl>

<bibl id="B88">
  <title><p>Assortative mixing in networks</p></title>
  <aug>
    <au><snm>Newman</snm><fnm>M. E. J.</fnm></au>
  </aug>
  <source>Physical Review Letters</source>
  <pubdate>2002</pubdate>
  <volume>89</volume>
  <fpage>208701</fpage>
</bibl>

<bibl id="B89">
  <title><p>Mixing patterns in networks</p></title>
  <aug>
    <au><snm>Newman</snm><fnm>M. E. J.</fnm></au>
  </aug>
  <source>Physical Review E</source>
  <pubdate>2003</pubdate>
  <volume>67</volume>
  <fpage>026126</fpage>
</bibl>

<bibl id="B90">
  <title><p>Condom use in multiethnic neighborhoods of {San Francisco} -- the
  population-based {AMEN (AIDS in Multiethnic Neighborhoods)} study</p></title>
  <aug>
    <au><snm>Catania</snm><fnm>J. A.</fnm></au>
    <au><snm>Coates</snm><fnm>T. J.</fnm></au>
    <au><snm>Kegels</snm><fnm>S.</fnm></au>
    <au><snm>Fullilove</snm><fnm>M. T.</fnm></au>
    <au><snm>Peterson</snm><fnm>J.</fnm></au>
    <au><snm>Marin</snm><fnm>B.</fnm></au>
    <au><snm>Siegel</snm><fnm>D.</fnm></au>
    <au><snm>Hulley</snm><fnm>S.</fnm></au>
  </aug>
  <source>American Journal of Public Health</source>
  <pubdate>1992</pubdate>
  <volume>82</volume>
  <fpage>284</fpage>
  <lpage>287</lpage>
</bibl>

<bibl id="B91">
  <title><p>We've already fought this war. {Available at:
  }\url{http://www.southdadenewsleader.com/opinion/we-ve-already-fought-this-war/article_ac44ce12-83b7-11e7-9a4c-0f8cc879c501.html}</p></title>
  <aug>
    <au><snm>Curbelo</snm><fnm>C</fnm></au>
  </aug>
  <source>South Dade News Leader</source>
  <pubdate>17 August 2017</pubdate>
</bibl>

<bibl id="B92">
  <title><p>What drives media slant? {Evidence} from {U.S.} daily
  newspapers</p></title>
  <aug>
    <au><snm>Gentzkow</snm><fnm>M.</fnm></au>
    <au><snm>Shapiro</snm><fnm>J. M.</fnm></au>
  </aug>
  <source>Econometrica</source>
  <pubdate>2010</pubdate>
  <volume>78</volume>
  <issue>1</issue>
  <fpage>35</fpage>
  <lpage>71</lpage>
</bibl>

<bibl id="B93">
  <title><p>Trump is dividing the country, {U.S.} voters say 2-1, {Quinnipiac
  University} national poll finds; most trust media more than
  {President}</p></title>
  <aug>
    <au><cnm>{Quinnipiac University}</cnm></au>
  </aug>
  <pubdate>23 August 2017</pubdate>
  <note>Available at
  \url{https://poll.qu.edu/national/release-detail?ReleaseID=2482}</note>
</bibl>

<bibl id="B94">
  <title><p>Political polarization in the {American} public: {H}ow increasing
  ideological uniformity and partisan antipathy affect politics, compromise,
  and everyday life</p></title>
  <aug>
    <au><snm>Dimrock</snm><fnm>M.</fnm></au>
    <au><snm>Carroll</snm><fnm>D.</fnm></au>
  </aug>
  <pubdate>2014</pubdate>
  <note>Available at
  \url{http://www.people-press.org/2014/06/12/political-polarization-in-the-american-public/}</note>
</bibl>

<bibl id="B95">
  <title><p>Fear and loathing across party lines: {N}ew evidence on group
  polarization</p></title>
  <aug>
    <au><snm>Iyengar</snm><fnm>S.</fnm></au>
    <au><snm>Westwook</snm><fnm>S. J.</fnm></au>
  </aug>
  <source>American Journal of Political Science</source>
  <pubdate>2015</pubdate>
  <volume>59</volume>
  <fpage>690</fpage>
  <lpage>-707</lpage>
</bibl>

<bibl id="B96">
  <title><p>The triumph of polarized partisanship in 2016: {Donald Trump's}
  improbable victory</p></title>
  <aug>
    <au><snm>Jacobson</snm><fnm>G. C.</fnm></au>
  </aug>
  <source>Political Science Quarterly</source>
  <pubdate>2017</pubdate>
  <volume>132</volume>
  <issue>1</issue>
  <fpage>9</fpage>
  <lpage>41</lpage>
</bibl>

<bibl id="B97">
  <title><p>Detecting automation of {Twitter} accounts: {A}re you a human, bot,
  or cyborg?</p></title>
  <aug>
    <au><snm>Chu</snm><fnm>Z.</fnm></au>
    <au><snm>Gianvecchio</snm><fnm>S.</fnm></au>
    <au><snm>Wang</snm><fnm>H.</fnm></au>
    <au><snm>Jajodia</snm><fnm>S.</fnm></au>
  </aug>
  <source>IEEE Transactions on Dependable and Secure Computing</source>
  <pubdate>2012</pubdate>
  <volume>9</volume>
  <issue>6</issue>
  <fpage>811</fpage>
  <lpage>824</lpage>
</bibl>

<bibl id="B98">
  <title><p>Bot-hunter: {A} tiered approach to detecting &amp characterizing
  automated activity on {Twitter}</p></title>
  <aug>
    <au><snm>Beskow</snm><fnm>D. M.</fnm></au>
    <au><snm>Carley</snm><fnm>K. M.</fnm></au>
  </aug>
  <pubdate>2018</pubdate>
  <note>Conference paper. SBP-BRiMS: International Conference on Social
  Computing, Behavioral-Cultural Modeling and Prediction and Behavior
  Representation in Modeling and Simulation</note>
</bibl>

<bibl id="B99">
  <title><p>Social bots distort the 2016 {U. S. Presidential} election online
  discussion</p></title>
  <aug>
    <au><snm>Bessi</snm><fnm>A.</fnm></au>
    <au><snm>Ferrara</snm><fnm>E.</fnm></au>
  </aug>
  <source>First Monday</source>
  <pubdate>2016</pubdate>
  <volume>21</volume>
  <issue>11</issue>
</bibl>

<bibl id="B100">
  <title><p>Reverse engineering socialbot infiltration strategies in
  {Twitter}</p></title>
  <aug>
    <au><snm>Freitas</snm><fnm>C.</fnm></au>
    <au><snm>Benevenuto</snm><fnm>F.</fnm></au>
    <au><snm>Ghosh</snm><fnm>S.</fnm></au>
    <au><snm>Veloso</snm><fnm>A.</fnm></au>
  </aug>
  <source>Proceedings of the 2015 IEEE/ACM International Conference on Advances
  in Social Networks Analysis and Mining 2015</source>
  <pubdate>2015</pubdate>
  <fpage>25</fpage>
  <lpage>32</lpage>
</bibl>

<bibl id="B101">
  <title><p>Bots increase exposure to negative and inflammatory content in
  online social systems</p></title>
  <aug>
    <au><snm>Stella</snm><fnm>M</fnm></au>
    <au><snm>Ferrara</snm><fnm>E</fnm></au>
    <au><snm>De Domenico</snm><fnm>M</fnm></au>
  </aug>
  <source>Proceedings of the National Academy of Sciences of the United States
  of America</source>
  <publisher>National Academy of Sciences</publisher>
  <pubdate>2018</pubdate>
  <volume>115</volume>
  <issue>49</issue>
  <fpage>12435</fpage>
  <lpage>-12440</lpage>
</bibl>

<bibl id="B102">
  <title><p>Detecting spam in a {Twitter} network</p></title>
  <aug>
    <au><snm>Yardi</snm><fnm>S.</fnm></au>
    <au><snm>Romero</snm><fnm>D.</fnm></au>
    <au><snm>Schoenebeck</snm><fnm>G.</fnm></au>
    <au><snm>Boyd</snm><fnm>D.</fnm></au>
  </aug>
  <source>First Monday</source>
  <pubdate>2010</pubdate>
  <volume>15</volume>
  <issue>1</issue>
</bibl>

<bibl id="B103">
  <title><p>{SocksCatch:} automatic detection and grouping of sockpuppets in
  social media</p></title>
  <aug>
    <au><snm>Yamak</snm><fnm>Z.</fnm></au>
    <au><snm>Saunier</snm><fnm>J.</fnm></au>
    <au><snm>Vercouter</snm><fnm>L.</fnm></au>
  </aug>
  <source>Knowledge-Based Systems</source>
  <pubdate>2018</pubdate>
  <volume>149</volume>
  <fpage>124</fpage>
  <lpage>142</lpage>
</bibl>

<bibl id="B104">
  <title><p>{HPSCI Minority Exhibit B}</p></title>
  <aug>
    <au><cnm>{House Permanent Select Committee on Intelligence Minority
  Staff}</cnm></au>
  </aug>
  <source>\url{https://democrats-intelligence.house.gov/uploadedfiles/exhibit_b.pdf}</source>
  <pubdate>1 November 2017</pubdate>
  <note>Available at
  \url{https://democrats-intelligence.house.gov/uploadedfiles/exhibit_b.pdf}</note>
</bibl>

<bibl id="B105">
  <title><p>{United States of America} v. {Internet Research Agency LLC},
  {Case} {1:18-cr-00032-DLF}. {18 U.S.C.} \textsection \textsection {2, 371,
  1349, 1028A}. {Available at:}
  \url{https://www.justice.gov/file/1035477/}.</p></title>
  <aug>
    <au><cnm>{United States District Court for the District of
  Columbia}</cnm></au>
  </aug>
  <pubdate>2018.</pubdate>
</bibl>

<bibl id="B106">
  <title><p>Voteview</p></title>
  <aug>
    <au><snm>Lewis</snm><fnm>J. B.</fnm></au>
    <au><snm>Poole</snm><fnm>K. T.</fnm></au>
    <au><snm>Rosenthal</snm><fnm>H.</fnm></au>
  </aug>
  <pubdate>2019</pubdate>
  <note>Available at \url{https://voteview.com}</note>
</bibl>

<bibl id="B107">
  <title><p>Multilayer networks</p></title>
  <aug>
    <au><snm>Kivel{\"a}</snm><fnm>M.</fnm></au>
    <au><snm>Arenas</snm><fnm>A.</fnm></au>
    <au><snm>Barthelemy</snm><fnm>M.</fnm></au>
    <au><snm>Gleeson</snm><fnm>J. P.</fnm></au>
    <au><snm>Moreno</snm><fnm>Y.</fnm></au>
    <au><snm>Porter</snm><fnm>M. A.</fnm></au>
  </aug>
  <source>Journal of Complex Networks</source>
  <pubdate>2014</pubdate>
  <volume>2</volume>
  <fpage>203</fpage>
  <lpage>-271</lpage>
</bibl>

<bibl id="B108">
  <title><p>On the Origins of Memes by Means of Fringe Web
  Communities</p></title>
  <aug>
    <au><snm>Zannettou</snm><fnm>S</fnm></au>
    <au><snm>Caulfield</snm><fnm>T</fnm></au>
    <au><snm>Blackburn</snm><fnm>J</fnm></au>
    <au><snm>De Cristofaro</snm><fnm>E</fnm></au>
    <au><snm>Sirivianos</snm><fnm>M</fnm></au>
    <au><snm>Stringhini</snm><fnm>G</fnm></au>
    <au><snm>Suarez Tangil</snm><fnm>G</fnm></au>
  </aug>
  <source>Proceedings of the Internet Measurement Conference 2018</source>
  <publisher>New York, NY, USA: ACM</publisher>
  <series><title><p>IMC '18</p></title></series>
  <pubdate>2018</pubdate>
  <fpage>188</fpage>
  <lpage>-202</lpage>
</bibl>

<bibl id="B109">
  <title><p>Who Let The Trolls Out?: Towards Understanding State-Sponsored
  Trolls</p></title>
  <aug>
    <au><snm>Zannettou</snm><fnm>S</fnm></au>
    <au><snm>Caulfield</snm><fnm>T</fnm></au>
    <au><snm>Setzer</snm><fnm>W</fnm></au>
    <au><snm>Sirivianos</snm><fnm>M</fnm></au>
    <au><snm>Stringhini</snm><fnm>G</fnm></au>
    <au><snm>Blackburn</snm><fnm>J</fnm></au>
  </aug>
  <source>Proceedings of the 10th ACM Conference on Web Science</source>
  <publisher>New York, NY, USA: ACM</publisher>
  <series><title><p>WebSci '19</p></title></series>
  <pubdate>2019</pubdate>
  <fpage>353</fpage>
  <lpage>-362</lpage>
</bibl>

<bibl id="B110">
  <title><p>A quantitative approach to understanding online
  antisemitism</p></title>
  <aug>
    <au><snm>Finkelstein</snm><fnm>J.</fnm></au>
    <au><snm>Zannettou</snm><fnm>S.</fnm></au>
    <au><snm>Bradlyn</snm><fnm>B.</fnm></au>
    <au><snm>Blackburn</snm><fnm>J.</fnm></au>
  </aug>
  <source>arXiv</source>
  <pubdate>2018</pubdate>
  <volume>1809.10644</volume>
</bibl>

</refgrp>
} 


\end{document}